\documentclass{aa}
\usepackage{txfonts}
\usepackage{graphicx}
\usepackage{epsfig}

\usepackage{natbib}
\bibpunct{(}{)}{;}{a}{}{,}
\newcommand{\refeq}[1]{Eq.\,\ref{#1}}
\newcommand{\reffig}[1]{Fig.\,\ref{#1}}

\newcommand{\refsec}[1]{Sect.\,\ref{#1}}

\begin{document}

\title{Reconciling observed GRB prompt spectra\\ with synchrotron radiation ?}
\author{Fr\'{e}d\'{e}ric Daigne\inst{1}\thanks{Institut Universitaire de France} \and {\v Z}eljka Bo{\v s}njak\inst{2,1} \and Guillaume Dubus\inst{3,1}}
\institute{Institut d'Astrophysique de Paris, UMR 7095
Universit\'{e} Pierre et Marie Curie -- CNRS, 
98bis boulevard Arago, 75014 Paris, France 
\and
AIM (UMR 7158 CEA/DSM-CNRS-Universit\'e Paris Diderot) Irfu/Service d'Astrophysique, Saclay, 91191 Gif-sur-Yvette Cedex, France
\and
Laboratoire d'Astrophysique de Grenoble, UMR 5571 Universit\'{e} Joseph
Fourier -- CNRS, BP 53, 38041 Grenoble, France}

\offprints{F. Daigne (\texttt{daigne@iap.fr})}

\date{}
\titlerunning{Reconciling observed GRB prompt spectra with synchrotron radiation ?}
\authorrunning{Daigne, Bo{\v s}njak \& Dubus}

\abstract  
{Gamma-ray burst emission in the prompt phase is often interpreted as synchrotron radiation from high-energy electrons accelerated in internal shocks. Fast synchrotron cooling of a power-law distribution of electrons leads to the prediction that the slope below the spectral peak has a photon index $\alpha=-3/2$ ($N(E)\propto E^{\alpha}$). However, this differs significantly from the observed median value $\alpha\approx -1$. This discrepancy has been used to argue against this scenario.}
{We quantify the influence of inverse Compton (IC) and adiabatic cooling on the low energy slope to understand whether these processes can reconcile the observed slopes with a synchrotron origin.}
{We use a time-dependent code developed to calculate the GRB prompt emission within the internal shock model. The code follows both the shock dynamics and electron energy losses  and can be used to generate lightcurves and spectra. We investigate the dependence of the low-energy slope on the parameters of the model and identify parameter regions that lead to values $\alpha>-3/2$. }
{Values of $\alpha$ between $-3/2$ and $-1$ are reached when electrons suffer IC losses in the Klein-Nishina regime. This does not necessarily imply a strong IC component in the \textit{Fermi}/LAT  range (GeV) because scatterings are only moderately efficient. Steep slopes require that a large fraction (10-30\%)  of the dissipated energy is given to a small fraction  ($\la$1\%)  of the electrons and that the magnetic field energy density fraction remains low ($\la 0.1$\%). Values of $\alpha$ up to $-2/3$ can be obtained with relatively high radiative efficiencies ($>$50\%) when adiabatic cooling is comparable with radiative cooling (marginally fast cooling). This requires collisions at small radii and/or with low magnetic fields.}
{Amending the standard fast cooling scenario to account for IC cooling naturally leads to values $\alpha$ up to $-1$. Marginally fast cooling may also account for values of $\alpha$ up to $-2/3$, although the conditions required are more difficult to reach. 
About 20~\% of GRBs show spectra with slopes $\alpha > -2/3$.
Other effects, not investigated here, such as a thermal component in the electron distribution or pair production by high-energy gamma-ray photons may further affect $\alpha$. 
Still, the majority of observed GRB prompt spectra can be reconciled with a synchrotron origin, constraining the microphysics of mildly relativistic internal shocks.
}

\maketitle

% ====================
% SECTION 1 : Introduction
% ====================

\section{Introduction}
The physical origin of the prompt emission in Gamma-Ray Bursts (hereafter GRBs) is still uncertain. The identification of the dominant energy reservoir in the relativistic outflow, of
the mechanism responsible for its extraction and of the processes by which the dissipated energy is eventually radiated remains a major unresolved issue.
There are three potential energy reservoirs : (i) thermal energy that can be radiated at the photosphere \citep{meszaros:00,daigne:02,giannios:07,peer:08,beloborodov:10}, (ii) kinetic energy that can be extracted by shock waves propagating within the outflow and then radiated by shock-accelerated electrons (internal shocks, \citet{rees:94,kobayashi:97,daigne:98}), or (iii) magnetic energy that can be dissipated via the reconnection of field lines \citep{thompson:94,meszaros:97,spruit:01,drenkhahn:02,lyutikov:03,giannios:05} and then radiated by accelerated particles. In the two last cases, 
the expected dominant radiative processes are synchrotron radiation and inverse Compton scattering.\\

Observed GRB spectra can provide reliable constraints on the extraction mechanism and dominant radiative process. 
A typical GRB prompt emission spectrum is usually well described by a phenomenological
model \citep{band:93} where the photon flux follows $N(E)\propto E^{\alpha}$ at low energies and $N(E)\propto E^{\beta}$ at high energies,
with a smooth
transition around $E_\mathrm{p}$, which is the peak energy of the $\nu
F_{\nu}$ spectrum. Typical values in GRBs observed by BATSE are $\alpha\sim -1$ and $\beta\sim -2.3$ for the low- and high-energy photon index and $E_\mathrm{p}\sim 250\, \mathrm{keV}$ for the peak energy \citep{preece:00}. 
It is well known that the observed value $\alpha\sim -1$ 
is in clear contradiction with the predicted value for the synchrotron radiation from relativistic electrons \citep{preece:98,ghisellini:00}.
One expects $\alpha=-2/3$ in slow cooling regime and $\alpha=-3/2$ in fast cooling regime \citep{sari:98}, the latter case being
favored in the prompt GRB phase as the slow cooling regime would lead both to an energy crisis and a difficulty to reproduce the shortest timescale variability \citep{rees:94,sari:96,kobayashi:97}. \\

On the other hand, synchrotron radiation is a very natural expectation for the emission from shock-accelerated electrons. In GRBs especially, it is most probably at work in afterglows. 
Observations of prompt GRBs by the LAT instrument on board \textit{Fermi} indicate that most GRBs do not show an additional component at high energy ($>100\, \mathrm{MeV}$) 
brighter or 
as bright as in the soft gamma-ray range \citep{abdo:09,omodei:09}. Prompt observations in the optical domain remain difficult but do not show strong evidence in favor of a bright additional component at low energy, with some notable exceptions like GRB 080319B \citep{racusin:08}.
This strongly favors synchrotron radiation compared to the synchrotron self-Compton (SSC) process  for the emission observed in the soft gamma-ray domain (keV--MeV). Indeed, the latter requires some fine tuning to avoid a strong component either in the optical-UV-soft X-rays domain, or in the GeV range \citep{bosnjak:09,zou:09,piran:09}. Compared to SSC, synchrotron radiation has also a better ability -- at least in the internal shock framework -- to reproduce the observed spectral evolution : e.g. hardness-intensity and hardness-fluence correlations, evolution of the pulse width with energy channel, time lags \citep[][{Bo{\v s}njak} 2010 in preparation]{daigne:98,ramirezruiz:00,daigne:03}.\\

In this paper, we investigate a solution to steepen the low-energy slope of the synchrotron component and possibly reconcile the synchrotron process with observed GRB prompt spectra. This solution is related to the steepening of the low-energy synchrotron slope by moderately efficient inverse Compton scatterings in Klein-Nishina regime, as suggested by \citet{derishev:01,bosnjak:09,nakar:09}; see also \citet{rees:67} where this is mentionned in the general context of SSC radiation. It is an alternative to the SSC scenario \citep[see e.g.][]{panaitescu:00,baring:04,kumar:08}, to the comptonization scenario \citep{liang:97,ghisellini:99}, or to other propositions to modify the standard synchrotron radiation, related to the timescale of the acceleration process \citep{stern:04,asano:09}, the pitch-angle distribution of electrons \citep{lloyd-ronning:02} or the small scale structure of the magnetic field \citep{medvedev:00,peer:06}. 
A summary of the measurements of the low-energy photon index $\alpha$ and its distribution in GRBs is given in \refsec{sec:catalog}.
Then
\refsec{sec:theory} describes how the standard synchrotron spectrum 
is affected by additional processes such as inverse Compton scatterings or adiabatic cooling. It allows to identify physical conditions -- in terms of intensity of the magnetic field, distribution of relativistic electrons, etc. -- that lead to low-energy slopes steeper than the standard prediction $\alpha=-3/2$. 
We discuss in 
\refsec{sec:internalshocks} how such conditions could be found in GRB outflows. We compute 
expected pulse lightcurves and spectral evolution in the framework of the internal shock model and show that steep slopes are indeed expected in a large region of the parameter space. 
We summarize our conclusions and discuss future possible developments in \refsec{sec:conclusion}.

%%%%%%%%%%%%%%%%%%%%%%%%%%%%%%%%%%%%%%%%%%%%%%%%%%%%%%%%%
% Figure 1
%
\begin{figure*}[!t]
\begin{center}
\begin{tabular}{ccc}
\includegraphics[width=0.32\textwidth]{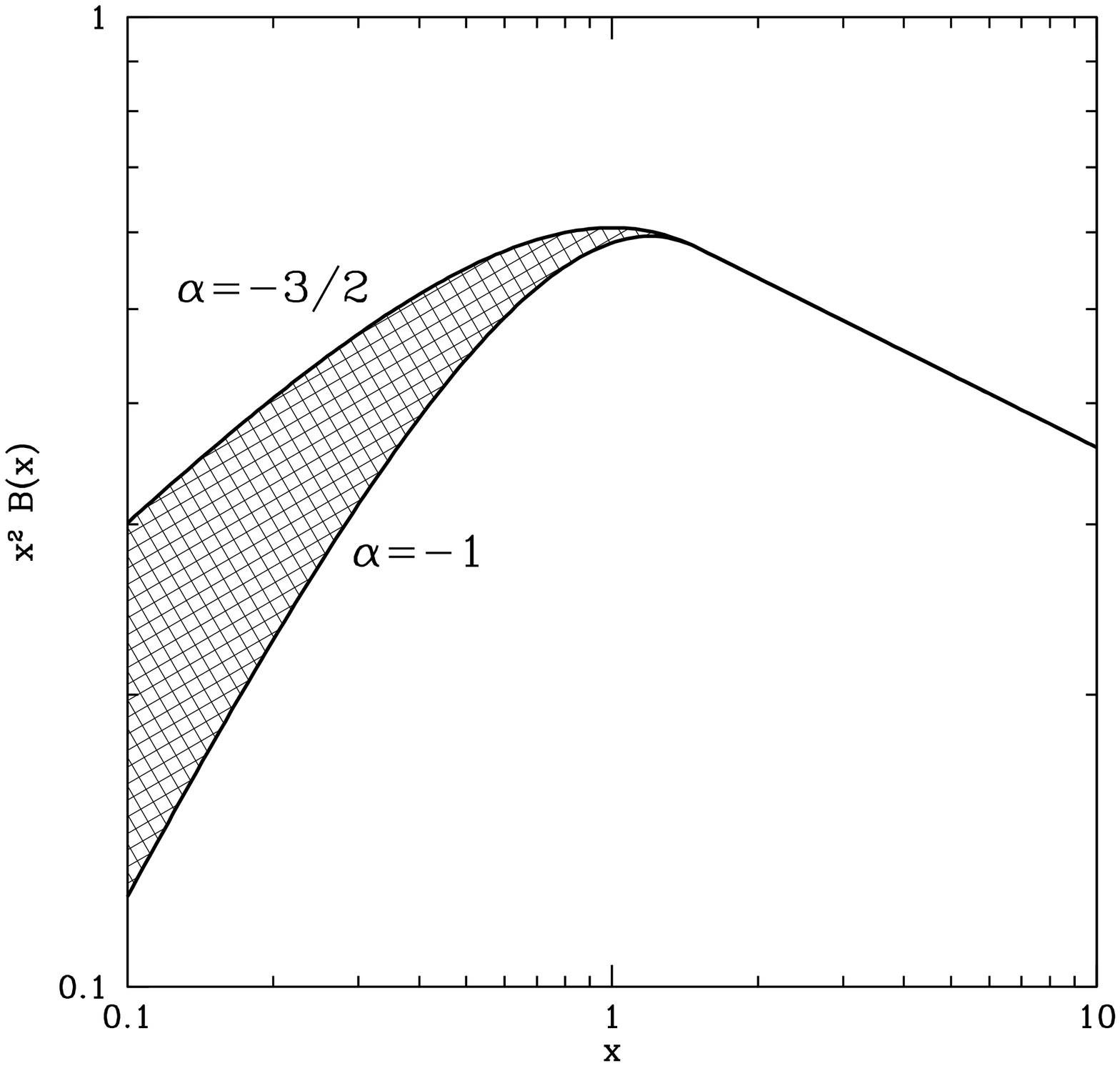} &
\includegraphics[width=0.32\textwidth]{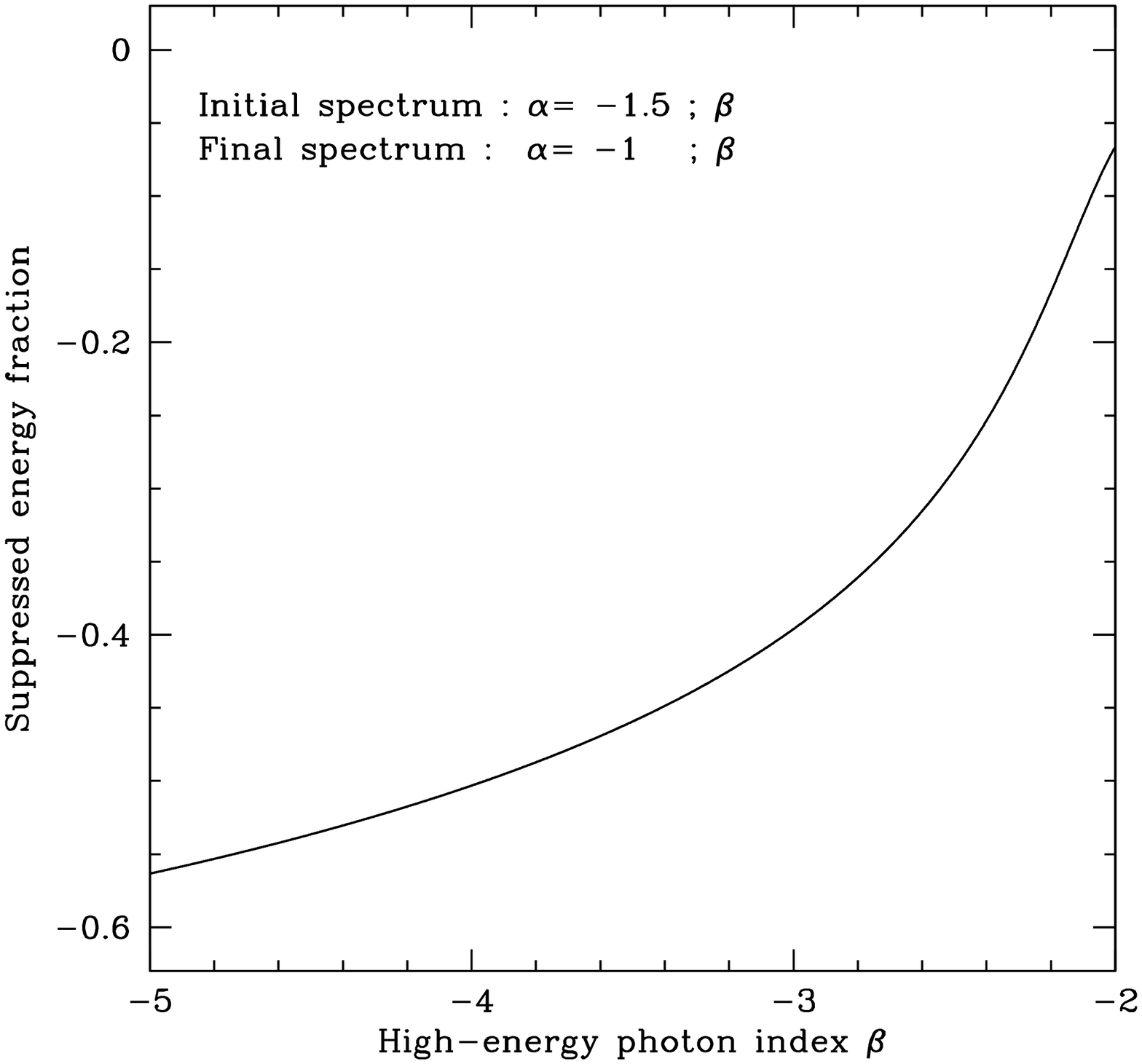} &
\includegraphics[width=0.32\textwidth]{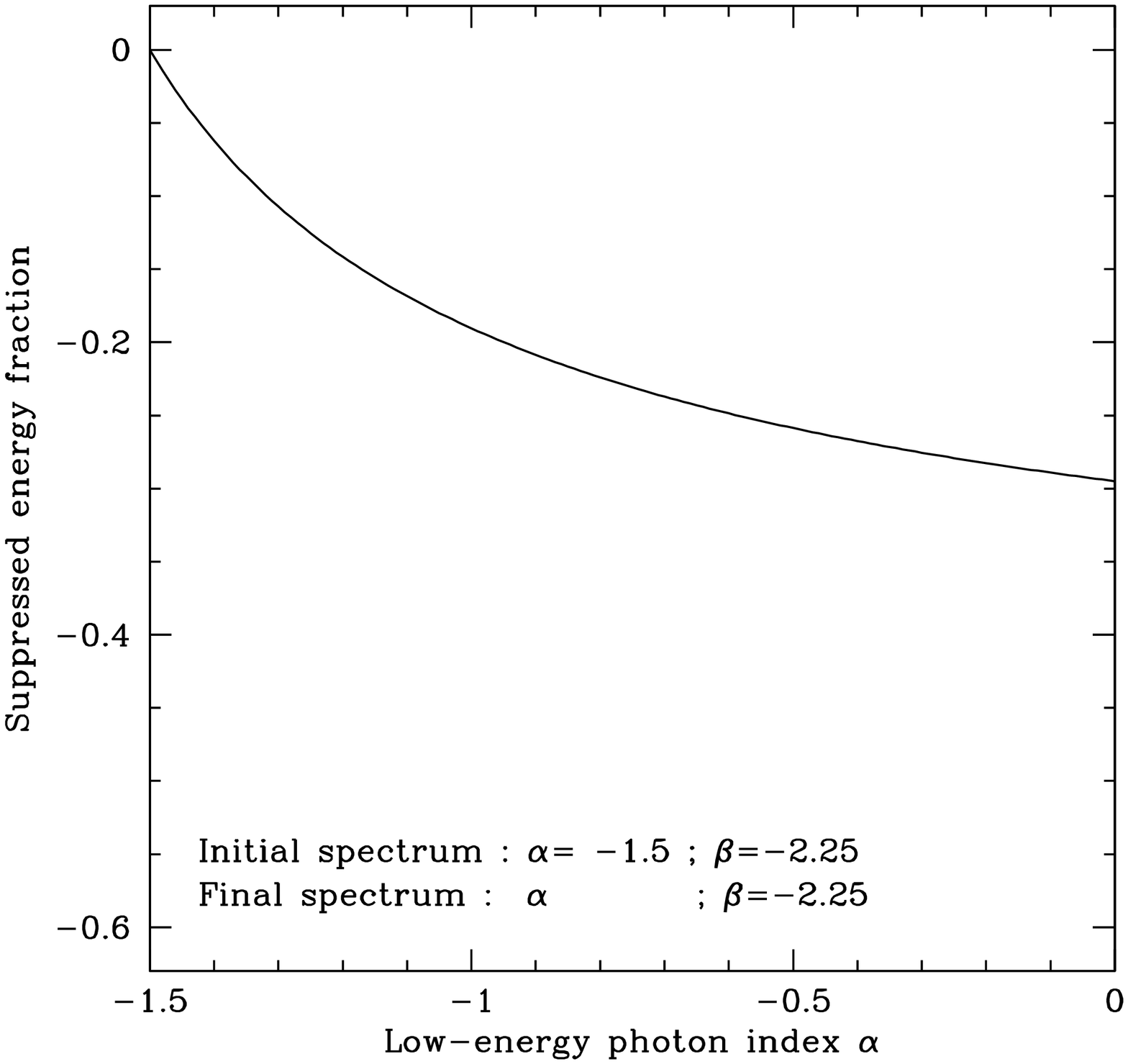} 
\end{tabular}
\end{center}
\vspace*{-3ex}

\caption{\textbf{Consequences on energetics of a steepening of the low-energy slope in the "Band" function.} \textit{Left:} the normalized "Band" function is plotted for $\beta=-2.25$ and $\alpha=-1.5$ (standard value for synchrotron radiation in fast cooling regime) or $\alpha=-1$ (mean value observed in GRB spectra). The dashed area corresponds to the fraction $f_\mathrm{e}$ of the energy that should be removed by any process leading to  such a change of the low-energy photon index. Here $f_\mathrm{e}\simeq 19\,\%$. \textit{Center:} the fraction $f_\mathrm{e}$ is now plotted as a function of $\beta$, assuming a low-energy photon index changing from $\alpha=-1.5$ to $1$. \textit{Right:} the fraction $f_\mathrm{e}$ is now plotted as a function of  $\alpha$, when the photon index is changing from $\alpha=-1.5$ to $\alpha$. We assume $\beta=-2.25$. These three figures are plotted assuming that the process responsible for the change of the slope $\alpha$ does not affect the value of the break energy $E_\mathrm{b}$ nor the high-energy tail of the spectrum.   
}
\label{fig:band}
\end{figure*}
%%%%%%%%%%%%%%%%%%%%%%%%%%%%%%%%%%%%%%%%%%%%%%%%%%%%%%%%%

% =====================
% SECTION 2 : Observations
% =====================

\section{A critical view on the observed distribution of GRB prompt spectral properties}
\label{sec:catalog}
We examine here in detail the observed distribution of the low energy
photon spectrum index $\alpha$, as it provides a relevant criteria for
the goodness of the emission model for GRB emission. To date the largest
database of gamma-ray burst high time and energy resolution data was
provided by the {Burst and Transient Source Experiment (BATSE)}
($\sim$20 keV - 2 MeV) on board the \textit{Compton Gamma Ray Observatory}.
\citet{kaneko:06} presented a systematic spectral analysis of 8459
time-resolved spectra from 350 GRBs (including 17 short events) observed
by BATSE; this sample includes also gamma--ray bursts that were
examined in previously published catalogs of BATSE GRBs \citep[e.g.][]{preece:98,preece:00}.  The reported distribution of $\alpha$ is
apparently not consistent with the predictions of the simple synchrotron
model: \citet{kaneko:06} showed that the median value for the time
resolved spectra $\alpha$ = --1.02$^{+0.26}_{-0.28}$ (long GRBs) and --0.87$^{+0.16}_{-0.39}$ (short GRBs).
The values obtained for the time integrated spectra indicate somewhat
softer spectra, with the respective median indices --1.15$^{+0.20}_{-0.22}$ and
--0.99$^{+0.21}_{-0.24}$ for long and short events respectively. The slope $\alpha$ is distributed roughly symmetrically around the median value.

The results obtained
by other instruments are consistent with BATSE observations: \citet{krimm:09}
combined the \textit{Swift} Burst Alert Telescope (BAT) and \textit{Suzaku} Wide band All-Sky
Monitor (WAM) data covering the broad energy band 15 to 5000 keV and report the distribution of $\alpha$ skewed toward slightly lower values, --1.23$\pm$0.28 (for time-integrated spectra), while \citet{pelangeon:08} for time integrated spectra of GRBs observed by \textit{High Energy Transient Explorer 2 (HETE-2)} in the energy band 2-400 keV find $\alpha$ = --1.08$\pm$0.20.
Similar results have been found recently by the
 \textit{Fermi Gamma--ray
Space Telescope} GBM and LAT in the broad energy range $\sim$8 keV to $>$ 100 GeV,
e.g. the sample studied by \citet{ghirlanda:10} of 12 GRBs
observed by \textit{Fermi} displays various values for $\alpha$, ranging from --1.26$\pm$0.04 (GRB 090618) to --0.55$\pm$0.07 (GRB 081222) for time-integrated spectra with determined peak energy. 

The observed distributions of GRB spectral parameters should however be
considered with precaution. We point out the possible caveats for interpreting the results of spectral analysis:
\begin{enumerate}

\item {the spectral analysis are commonly performed on bright
GRBs (with higher photon flux), which tend to have higher peak energies in general than dim GRBs \citep{mallozzi:95,borgonovo:01}. For that reason the spectra with higher $E_{peak}$ may be oversampled. 
The results of the time-resolved spectral analysis may be biased in the similar way: as the data were sampled more frequently during the intense episodes, the brighter portions of each burst may have more impact in the final distribution of spectral parameters \citep{kaneko:06};} 
\item {the low-energy photon spectra indices tend to correlate with the
peak energy of the $\nu F_{\nu}$ spectrum, the slope $\alpha$ becoming softer when the peak energy is decreasing \citep{kaneko:06,crider:97,lloyd-ronning:02,ford:95,preece:98}.
This effect might be due to a combination of the curvature of the spectrum around the peak energy and the limited spectral energy range sampled by the instruments;}
\item {as discussed by \citet{preece:98} and \citet{lloyd:00} for 
the BATSE spectra, the data don't always approach the GRB spectral low energy power law within the instrument energy range.  If peak energy is close to the edge of the
instrumental energy window, the low energy spectral power law may not have reached yet its asymptotic value. In that case  lower values of $\alpha$ are determined (i.e. softer spectra). 
\citet{kaneko:06} attempted to account for this effect and applied as a better
measure of the actual low energy behavior the effective index
$\alpha_{eff}$ for BATSE data, defined as the tangential slope of the spectrum at 25 keV (the lower energy limit of the BATSE window);} 
\item {the low-energy spectral index distributions of time-integrated and
of time-resolved spectra are slightly different; it is expected due to the
evolution of the spectral parameters during the integration time. The
sharp spectral breaks are smeared over and the indices of time
integrated spectra appear softer than in the case of the time-resolved
spectra \citep{kaneko:06}. These two last points could imply that the true low-energy spectral slopes in prompt GRBs are even steeper than the observed median value.} 
\end{enumerate}

Another important aspect to examine in the observed distributions of
spectral indices concerns the contribution of an individual GRB to the
overall distribution. Since the brighter and longer events in general
contribute with larger number of spectra in the BATSE spectroscopic catalog, 
we have computed distributions of spectral parameters where each time bin of a given GRB is weighted by the corresponding fraction of the total fluence of the burst. 
In this way GRBs with different number of time resolved spectra in their time histories 
 have the same impact on the overall distribution. Using
the data by \citet{kaneko:06}, we find that:
\begin{itemize}
\item Only 5\% of GRBs have more than 50\% of their spectra with very soft low
energy slope, $\alpha <$ --1.5. Such slopes are probably related to the spectral curvature around the peak energy and do not raise a problem for the standard synchrotron scenario;
\item 70\% of GRBs in the sample have more than 50\% of their spectra with
a low energy photon index within the limits of the synchrotron model,
$-3/2 \le \alpha \le -2/3$ and 35\% of GRBs have more than 50\% of their spectra with $\alpha$ in the range $-3/2 \le \alpha \le -1$;
\item Only 20\% of GRBs in the sample have more than 50\% of their spectra with
$\alpha >-2/3$. However most of these spectra have $\alpha$ close to $-2/3$. For instance, only 5\% of GRBs in the sample have more than 50\% of
spectra with $\alpha > -0.3$.
\end{itemize}
From these values, it appears that the synchrotron radiation should in principle be compatible with most observed prompt GRB spectra. One should however understand why the mean value of $\alpha$ is steeper than the expected slope for the fast cooling regime, $\alpha=-3/2$. 

%%%%%%%%%%%%%%%%%%%%%%%%%%%%%%%%%%%%%%%%%%%%%%%%%%%%%%%%%
% Figure 2
%
\begin{figure*}[!ht]
\begin{center}
\begin{tabular}{cc}
\includegraphics[width=0.47\textwidth]{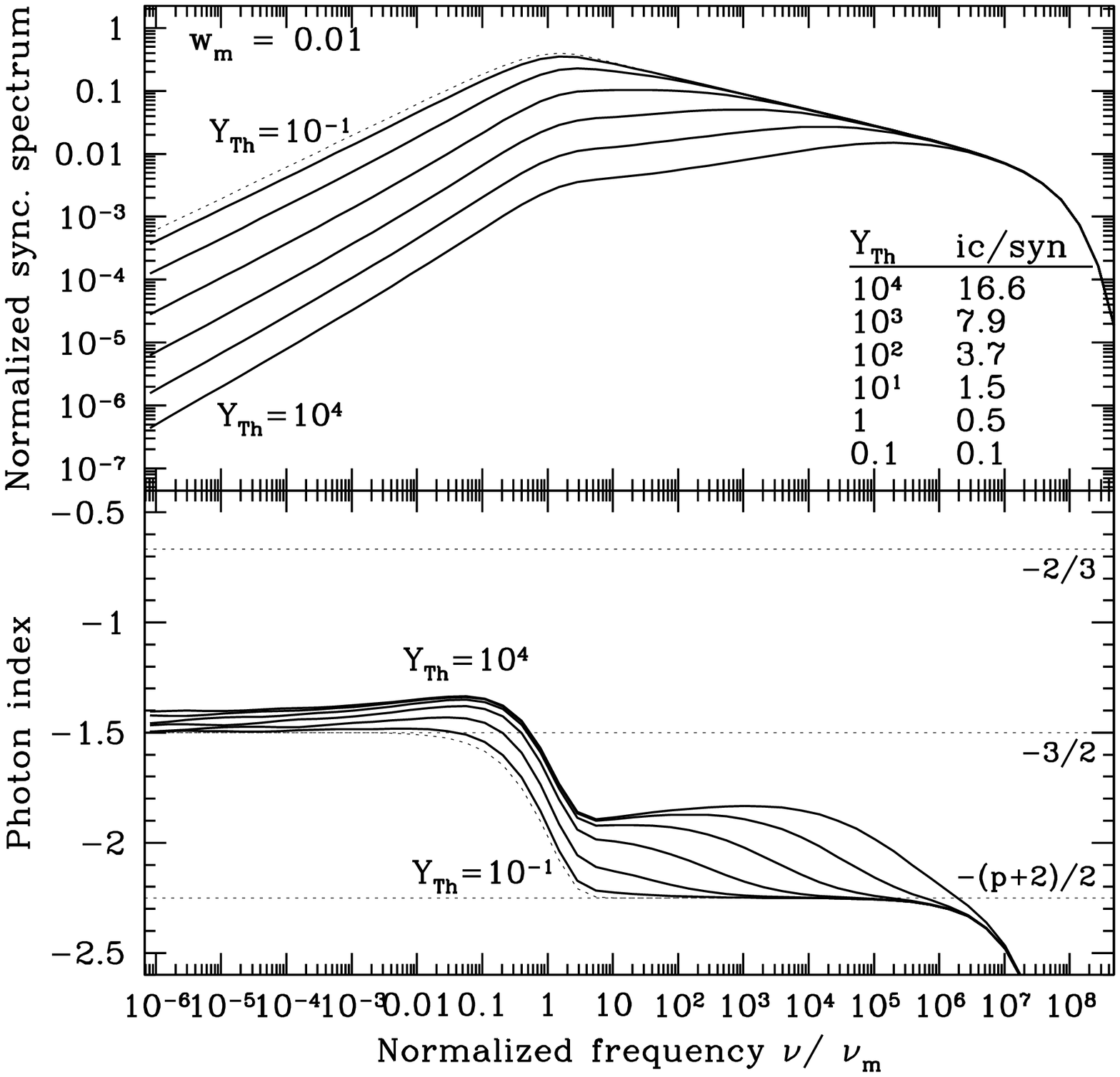} & \includegraphics[width=0.47\textwidth]{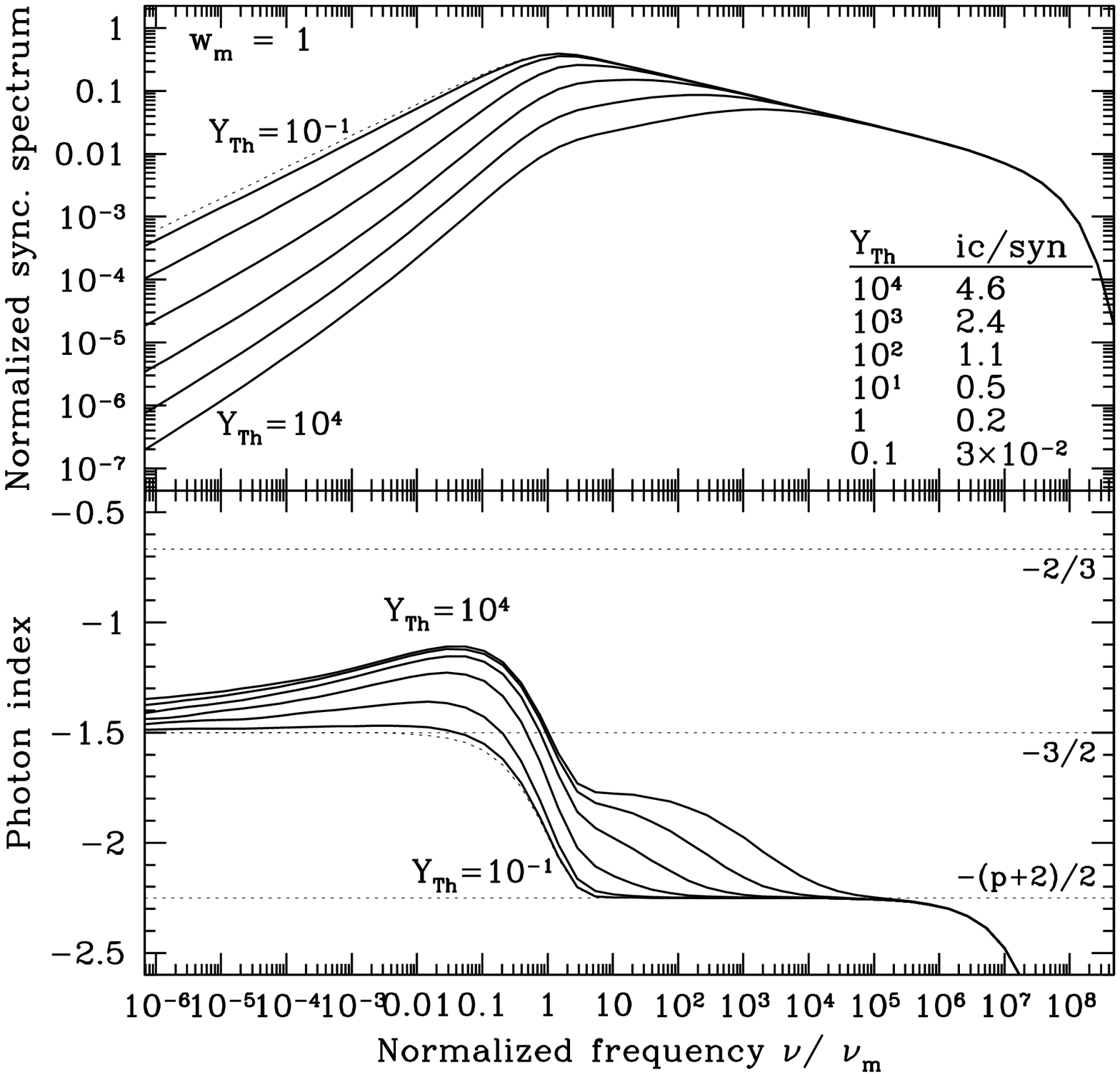} \\
\includegraphics[width=0.47\textwidth]{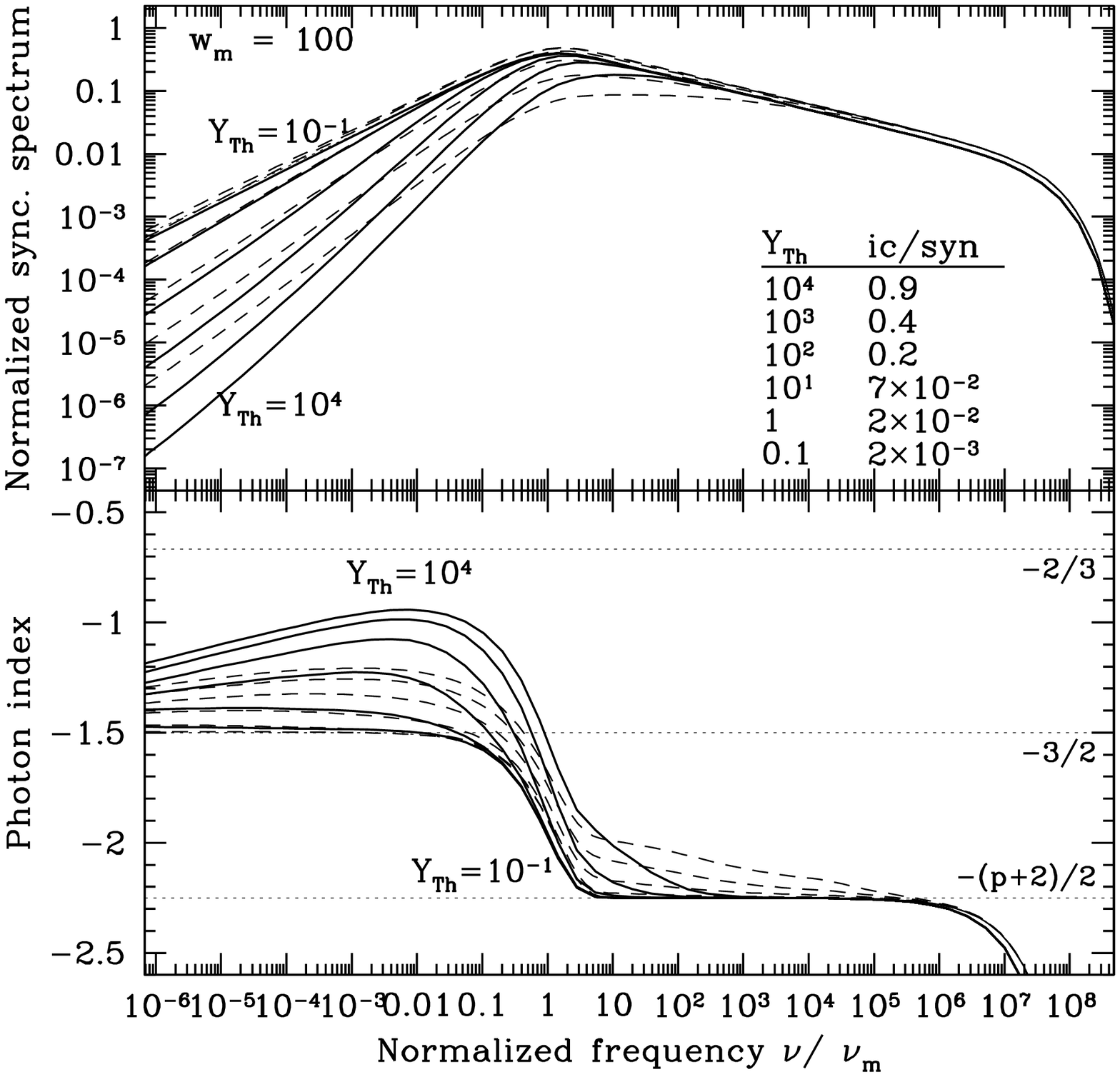} & \includegraphics[width=0.47\textwidth]{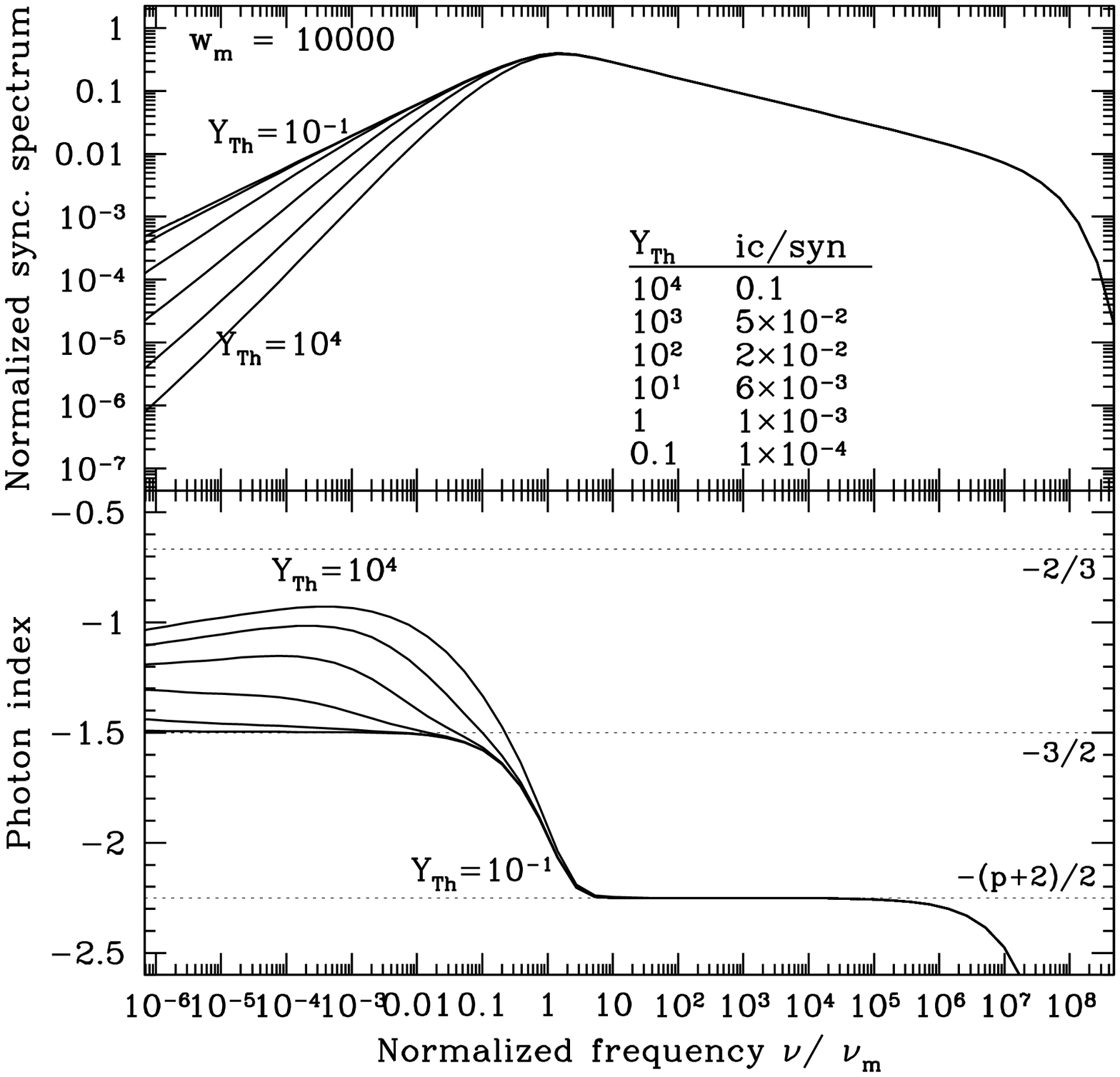} 
\end{tabular}
\end{center}
\caption{\textbf{The effect of inverse Compton scatterings in Klein-Nishina regime on the fast cooling synchrotron spectrum.} The normalized synchrotron spectrum defined by $\nu \left.u_{\nu}\right|_\mathrm{syn}/ n_\mathrm{e}^\mathrm{acc}\Gamma_\mathrm{m}m_\mathrm{e}c^2$ is plotted as a function of the normalized frequency $\nu/\nu_\mathrm{m}$, as well as the corresponding photon index $d\ln{u_{\nu}}/d\ln{\nu}-1$.
All spectra in thick solid line are computed numerically using a detailed radiative code including synchrotron radiation and inverse Compton scatterings (see text). Spectra in thin dotted line are computed with synchrotron radiation only. For clarity purposes, all other processes (adiabatic cooling, synchrotron self-absorption, $\gamma\gamma$ annihilation) are neglected.
A ratio $\Gamma_\mathrm{c}/\Gamma_\mathrm{m}=10^{-6}$ is assumed to ensure that all electrons are in fast cooling regime. The maximum Lorentz factor of electrons is fixed to $\Gamma_\mathrm{max}=10^4\Gamma_\mathrm{min}$ so the high energy cutoff in the synchrotron spectrum appears at the same frequency in all cases plotted here. The four panels correspond to increasing values of the $w_\mathrm{m}$ parameter, $w_\mathrm{m}=0.01$, $1$, $100$ and $10^4$ from the top-left to the bottom-right panel, i.e. to a growing importance of Klein-Nishina corrections for inverse Compton scatterings. In each panel, the six curves in solid line correspond to increasing values of the $Y$ parameter, $Y=0.1$, $1$, $10$, $10^2$, $10^3$ and $10^4$, i.e. to a growing efficiency of inverse Compton scatterings. The table inserted in each panel lists the values of the ratio $\mathcal{E}_\mathrm{ic}/\mathcal{E}_\mathrm{syn}$ of the inverse Compton component (not plotted here) over the  synchrotron component. In the bottom-left panel ($w_\mathrm{m}=100$), the synchrotron spectrum obtained assuming  a slow injection over $t_\mathrm{ex}$ (see text) is also plotted in dashed line for comparison.}
\label{fig:effectKN}
\end{figure*}
%%%%%%%%%%%%%%%%%%%%%%%%%%%%%%%%%%%%%%%%%%%%%%%%%%%%%%%%%

% =============================================
% SECTION 3 : Synchrotron radiation in fast cooling regime
% =============================================

\section{Synchrotron radiation in radiatively efficient regime}
\label{sec:theory}
In this section, all quantities are given in the comoving frame of the emitting region.

% SECTION 3.1 : The standard prediction
% --------------------------------------------------

\subsection{The standard prediction in fast cooling regime}
The synchrotron power of an electron with Lorentz factor $\gamma$ is given  by
\begin{equation}
P_\mathrm{syn}(\gamma) = \frac{\sigma_\mathrm{T}c}{6\pi} B^2 \gamma ^2\, 
\end{equation}
where $B$ is the magnetic field. If the source is relativistically expanding, as expected in gamma-ray bursts, adiabatic cooling competes with radiation.
This cooling process occurs on a typical dynamical timescale $t_\mathrm{ex}$. It is then convenient to define $\Gamma_\mathrm{c}$ as the Lorentz factor of electrons whose synchrotron radiative timescale equals $t_\mathrm{ex}$ \citep{sari:98} :
\begin{equation}
\Gamma_\mathrm{c} = \frac{6\pi m_\mathrm{e}c}{\sigma_\mathrm{T} B^2 t_\mathrm{ex}}\ .
\label{eq:gc}
\end{equation}  
If the GRB prompt emission comes from the radiation of relativistic electrons, it is necessary, both to allow for the shortest timescales observed in GRB lightcurves and to minimize the constraint on the energy budget, that these electrons are radiatively efficient, i.e. that their radiative timescale is shorter than $t_\mathrm{ex}$ \citep{rees:94,sari:96,kobayashi:97}. When only synchrotron radiation is considered, this is equivalent to the condition that most injected electrons by the acceleration process must have $\gamma > \Gamma_\mathrm{c}$. The resulting synchrotron spectrum has been described by \citet{sari:98} when the initial distribution of relativistic electrons is a power-law of slope $-p$ above a minimum Lorentz factor $\Gamma_\mathrm{m}$. If synchrotron self-absorption is neglected, three asymptotic branches are predicted 
\begin{equation}
\frac{\nu \left. u_{\nu}\right|_\mathrm{syn}}{n_\mathrm{e}^\mathrm{acc} \Gamma_\mathrm{m}m_\mathrm{e}c^2}\simeq
\left\lbrace\begin{array}{cl}
\left(\frac{\nu_\mathrm{c}}{\nu_\mathrm{m}}\right)^{1/2}\left(\frac{\nu}{\nu_\mathrm{c}}\right)^{4/3} & \mathrm{if}\ \nu < \nu_\mathrm{c}\\
\left(\frac{\nu}{\nu_\mathrm{m}}\right)^{1/2} & \mathrm{if}\ \nu_\mathrm{c} \le \nu \le \nu_\mathrm{m}\\
\left(\frac{\nu}{\nu_\mathrm{m}}\right)^{-\frac{p-2}{2}} & \mathrm{if}\ \nu > \nu_\mathrm{m}\\
\end{array}\right.\ ,
\end{equation}
where $u_{\nu}$ is the final photon energy density at frequency $\nu$ and $n_\mathrm{e}^\mathrm{acc}$ is the initial density of relativistic electrons. The break frequency $\nu_\mathrm{m}$ (resp. $\nu_\mathrm{c}$) is defined as the synchrotron frequency for an electron with Lorentz factor $\Gamma_\mathrm{m}$ (resp. $\Gamma_\mathrm{c}$). This asymptotic spectrum shows clearly that the predicted photon spectral slope $\alpha$ below the peak of $\nu F_{\nu}$ is $-3/2$, in apparent contradition with observations.

% SECTION 3.2 : The effect of Inverse Compton scatterings in Klein-Nishina regime
% -------------------------------------------------------------------------------------------------------

\subsection{The effect of Inverse Compton scatterings in Klein-Nishina regime}
The mean value of $\alpha$ observed in the BATSE spectroscopic catalog (see \refsec{sec:catalog}) is close to $-1$. As seen in \reffig{fig:band}, for typical values of the high-energy photon index $\beta$ between $-2$ and $-3$, a change of the low-energy slope of the Band function from $\alpha=-1.5$ to $-1$ requires only a sub-dominant radiative process that can transfer about $20-40\,\%$ of the energy from the synchrotron component into another component. A natural candidate is  inverse Compton scattering of synchrotron photons by relativistic electrons. Indeed, this process necessarily takes place in the emitting region. Two parameters can be introduced to evaluate the importance of inverse Compton scatterings. The first parameter, $w_\mathrm{m}$, measures if scatterings occur mostly in Thomson regime ($w_\mathrm{m}\ll 1$) or if Klein-Nishina corrections are important. It is defined by
\begin{equation}
w_\mathrm{m} = \Gamma_\mathrm{m} \epsilon_\mathrm{m}\ .
\label{eq:wm}
\end{equation}
The second parameter, $Y_\mathrm{Th}$, measures the intensity of the inverse Compton process in the Thomson regime and is defined by 
\begin{equation}
Y_\mathrm{Th} =  \sigma_\mathrm{T} n_\mathrm{e}^\mathrm{acc} c t_\mathrm{syn}\left(\Gamma_\mathrm{m}\right)  \times \frac{4}{3} \Gamma_\mathrm{m}^2\ ,
\label{eq:yth}
\end{equation}
where 
\begin{equation}
t_\mathrm{syn}\left(\gamma\right) = \frac{\gamma m_\mathrm{e} c^2}{P_\mathrm{syn}(\gamma)}= t_\mathrm{ex} \frac{\Gamma_\mathrm{c}}{\gamma}
\end{equation} 
is the synchrotron timescale of an electron with Lorentz factor $\gamma$. In \refeq{eq:yth} the first term, $\sigma_\mathrm{T} n_\mathrm{e}^\mathrm{acc} c t_\mathrm{syn}$, is the Thomson optical depth associated with relativistic electrons in fast cooling regime, and the second term, $\frac{4}{3} \Gamma_\mathrm{m}^2$ , corresponds to the typical boost of a photon scattered by an electron at Lorentz factor $\Gamma_\mathrm{m}$, if the scattering occurs in Thomson regime ($w_\mathrm{m}\ll 1$). Therefore, if $w_\mathrm{m}\ll 1$ and $Y_\mathrm{Th}\ll 1$, the ratio of the total energy $\mathcal{E}_\mathrm{ic}$ radiated in the inverse Compton component  over the total energy $\mathcal{E}_\mathrm{syn}$ radiated in the synchrotron component is simply given by $\mathcal{E}_\mathrm{ic}/\mathcal{E}_\mathrm{syn}\simeq Y_\mathrm{Th}$. Still in the Thomson regime ($w_\mathrm{m}\ll 1$), if $Y_\mathrm{Th}$ is large, the effective radiative timescale becomes $\sim t_\mathrm{syn}/Y_\mathrm{Th}$ and $\mathcal{E}_\mathrm{ic}/\mathcal{E}_\mathrm{syn}\simeq \sqrt{Y_\mathrm{Th}}$. Finally, when Klein-Nishina corrections are important ($w_\mathrm{m}\ga 1$), both the cross section and the boost in frequency are significantly reduced so that $\mathcal{E}_\mathrm{ic}/\mathcal{E}_\mathrm{syn}\ll Y_\mathrm{Th}$ \citep[see][for details]{bosnjak:09}.\\

If the magnetic energy density represents a fraction $\epsilon_\mathrm{B}$ of the local energy density in the emitting region, and if the energy injected into accelerated relativistic electrons represents a fraction $\epsilon_\mathrm{e}$ of the same energy reservoir, \refeq{eq:yth} can be simplified to give 
\begin{equation}
Y_\mathrm{Th} =\frac{p-2}{p-1} \frac{\epsilon_\mathrm{e}}{\epsilon_\mathrm{B}}\ .
\end{equation}

When considering only synchrotron radiation and inverse Compton scatterings, the spectral shape of the synchrotron component, i.e. $\left.\nu\,u_{\nu}\right|_\mathrm{syn}/ n_\mathrm{e}^\mathrm{acc}\Gamma_\mathrm{m}m_\mathrm{e}c^2$ as a function of $\nu/\nu_\mathrm{m}$, depends only on these two parameters $w_\mathrm{m}$ and $Y_\mathrm{Th}$, in the limit of extreme fast cooling ($\Gamma_\mathrm{c}\ll \Gamma_\mathrm{m}$).  
It is well known that in Thomson regime ($w_\mathrm{m}\ll 1$), as the electron cooling rate due to inverse Compton scatterings remains proportional to $\gamma^2$ like for the synchrotron power, the spectral shape of the synchrotron component is un-affected by the scatterings \citep[see e.g.][]{sari:01,bosnjak:09}. To change the value of $\alpha$, the physical conditions in the emitting regions must necessarily be such that $w_\mathrm{m}\ga 1$. Then, Klein-Nishina corrections become important for most of the scatterings and the dependence of the electron cooling rate on $\gamma$ differs from $\gamma^2$. \citet{derishev:01} have shown that it results in a steeper slope $\alpha$, that can potentially reach the value $\alpha=-1$. We have shown in a previous paper that this change of the synchrotron slope due to inverse Compton scatterings in Klein-Nishina regime was indeed observed in detailed radiative calculations \citep{bosnjak:09}.  More recently, \citet{nakar:09} have presented a complete analytical estimate of the asymptotic synchrotron spectrum in the presence of inverse Compton scatterings, and have shown that the asymptotic slope $\alpha=-1$ is expected in the synchrotron component below the peak of $\nu F_\nu$ when\footnote{In \citet{nakar:09}, the situation where the slope $\alpha=-1$ is possible corresponds to cases IIb and IIc, which -- in addition to fast cooling -- are limited by $\left(\Gamma_\mathrm{m}/\hat{\Gamma}_\mathrm{m}\right)^{1/3}< \epsilon_\mathrm{e}/\epsilon_\mathrm{B}< \left(\Gamma_\mathrm{m}/\hat{\Gamma}_\mathrm{m}\right)^{3}$, where the authors define $\hat{\gamma}=m_\mathrm{e} c^2/ h \nu_\mathrm{syn}(\gamma)$ so that $w_\mathrm{m}=\Gamma_\mathrm{m}/\hat{\Gamma}_\mathrm{m}$, leading to the conditions given in \refeq{eq:condwmY}.} $\Gamma_\mathrm{c}\ll \Gamma_\mathrm{m}$ (fast cooling), $w_\mathrm{m}\gg 1$ (Klein-Nishina regime) and 
\begin{equation}
w_\mathrm{m}^{1/3} \le Y \le w_\mathrm{m}^3\ .
\label{eq:condwmY}
\end{equation}
%%%%%%%%%%%%%%%%%%%%%%%%%%%%%%%%%%%%%%%%%%%%%%%%%%%%%%%%%
% Figure 3
%
\begin{figure}[t!]
\centerline{\includegraphics[width=0.5\textwidth]{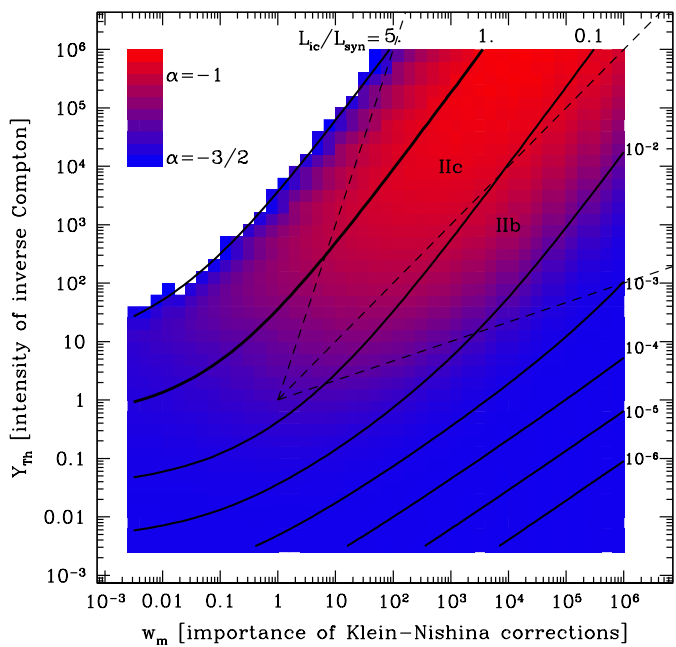}}
\caption{\textbf{The low-energy slope $\alpha$ of the fast cooling synchrotron spectrum in the presence of inverse Compton scatterings in Klein-Nishina regime.} In the $w_\mathrm{m}$--$Y_\mathrm{Th}$ plane, the value of the photon index $\alpha$ of the synchrotron spectrum below the peak of $\nu F_{\nu}$ is color-coded. All spectra have been computed numerically using a detailed radiative code including synchrotron radiation and inverse Compton scatterings (see text). Black thick lines of constant ratio $L_\mathrm{ic}/L_\mathrm{syn}$ are plotted on top of this diagram. Cases where $L_\mathrm{ic}/L_\mathrm{syn} > 5$ in the top-left corner are not plotted as we focus in this paper on situations where the synchrotron component is dominant.  In addition, thin dashed lines show the location of regions IIb and IIc from \citet{nakar:09}, where a slope $\alpha\to -1$ is predicted.}
\label{fig:diagKN}
\end{figure}
%%%%%%%%%%%%%%%%%%%%%%%%%%%%%%%%%%%%%%%%%%%%%%%%%%%%%%%%%
Therefore the analytical work of \citet{derishev:01} and \citet{nakar:09} and the numerical study of \citet{bosnjak:09} indicate that the effect of inverse Compton scatterings
in Klein-Nishina regime on the synchrotron component seems to be a promising possibility to explain observed values of $\alpha$ in the range $\left[-3/2;-1\right]$. However, as all possible values in this interval can be observed, it is necessary to go a step further than analytical estimates based on asymptotic behaviours. In \reffig{fig:effectKN}, we show the evolution of the synchrotron spectrum when $Y_\mathrm{Th}$ is increasing, for different values of $w_\mathrm{m}$ (and for $\Gamma_\mathrm{c}\ll \Gamma_\mathrm{m}$). The spectra are computed using the radiative code described in \citet{bosnjak:09} that solves simultaneously the equations of the time-evolution of the electron and photon distributions and that includes most relevant processes (adiabatic cooling, synchrotron radiation and self-absorption, inverse Compton scatterings and photon--photon annihilation). To focus on the effect described in this subsection, the spectra in \reffig{fig:effectKN} are computed including only synchrotron radiation and inverse Compton scatterings and do not take into account the other processes, whose impact is discussed later in the paper.  The slope $\alpha$ is clearly steepening continuously from $-3/2$ to $-1$. In \reffig{fig:diagKN}, we have color-coded in the diagram $w_\mathrm{m}$--$Y_\mathrm{Th}$ the value of the low-energy photon index of the synchrotron spectrum. This slope is only a representative value as the synchrotron spectrum below the peak shows some curvature with an evolving slope (see \reffig{fig:effectKN}). In practice, we adopt for $\alpha$ the clear maximum of the slope below $\nu_\mathrm{m}$.
If  
$\alpha=-3/2$ appears indeed universal when inverse Compton scatterings are negligible ($Y_\mathrm{Th}\ll 1$) or Klein-Nishina corrections unimportant ($w_\mathrm{m}\ll 1$), the situation is rather different in the quarter of the diagram where $w_\mathrm{m}\ge 1$ and $Y_\mathrm{Th}\ge 1$. In particular, a region where $\alpha$ evolves from $-3/2$ to $-1$ is well defined, whose boundaries agree reasonably well with the condition given by \refeq{eq:condwmY}.
In the same figure, lines of constant ratio $\mathcal{E}_\mathrm{ic}/\mathcal{E}_\mathrm{syn}$ are also plotted. In agreement with the analysis made at the beginning of this subsection, this ratio is typically in the range 0.1-1 in the region of interest (it is of course much smaller than $Y_\mathrm{Th}$ due to Klein-Nishina corrections to the cross section and the boost in frequency). 
It is important to note that $\alpha\simeq -1$ can be reached in practice for not too extreme values of the parameters, typically $w_\mathrm{m}=100-10^4$ and $Y\ga w_\mathrm{m}$. 

The maximum value of $\alpha$ that can be reached depends on the assumptions about the electron acceleration process. 
In principle, one would wish to follow the full 'magneto-hydrodynamical' evolution of the shocked region at the plasma scale. The framework used in the present study to follow the dynamics of shocks in a relativistic outflow limits us  to make simple assumptions on the electron injection timescale. Two extreme cases are possible.
If electrons are injected regularly over a timescale $t_\mathrm{injec}$ comparable to $t_\mathrm{ex}$ (as assumed in \citealt{nakar:09}), the slope will never be steeper than $\alpha=-1$ and in practice will be a little less steep than this absolute limit for reasonable choice of parameters (not too extreme $w_\mathrm{m}$ and $Y$ parameters, see \citealt{nakar:09} and bottom-left panel in \reffig{fig:effectKN}). If the injection occurs faster, the limit $\alpha=-1$ is more easily reached and even steeper values can be obtained for extreme sets of parameters (see the maximum slopes obtained for $Y_\mathrm{th}=10^4$ and $w_\mathrm{m}=100$ or $10000$ in \reffig{fig:effectKN}). Calculations presented in this paper corresponds to the regime where $t_\mathrm{injec}\ll t_\mathrm{ex}$.

% SECTION 3.3 : Additional effects
% ------------------------------------------

\subsection{Additional effects}
\subsubsection{Adiabatic cooling}

The process described in the previous subsection offers a physical interpretation of the observed values of the low-energy photon index in the range $[-3/2;-1]$. However the value $\alpha=-1$ does not appear as a limit in the observed distribution of $\alpha$ and an additional explanation has to be found for the steeper slopes.
When electrons are in slow cooling regime ($\Gamma_\mathrm{c}\gg \Gamma_\mathrm{m}$), the predicted value of the photon index below the peak of $\nu F_\mathrm{\nu}$ is $\alpha=-2/3$ \citep{sari:98}. Therefore, it is now tempting to associate spectra with $-1\le \alpha\le -2/3$ to this regime. There is however a problem due to the radiative efficiency $f_\mathrm{rad}=u_{\gamma}/u_\mathrm{e}^\mathrm{acc}$, where $u_\mathrm{e}^\mathrm{acc}$ is the initial energy density injected in relativistic electrons and $u_\mathrm{\gamma}$ the final energy density of the radiated photons. In the slow cooling regime $f_\mathrm{rad}$ is low, which increases the required energy budget to an uncomfortable level. Here, we rather consider the situation where electrons are in fast cooling regime but not deeply in this regime, i.e. $\Gamma_\mathrm{c}\la \Gamma_\mathrm{m}$ rather than $\Gamma_\mathrm{c}\ll\Gamma_\mathrm{m}$ ("marginally fast cooling regime").

\begin{figure*}[t]
\begin{center}
\begin{tabular}{cc}
\includegraphics[width=0.47\textwidth]{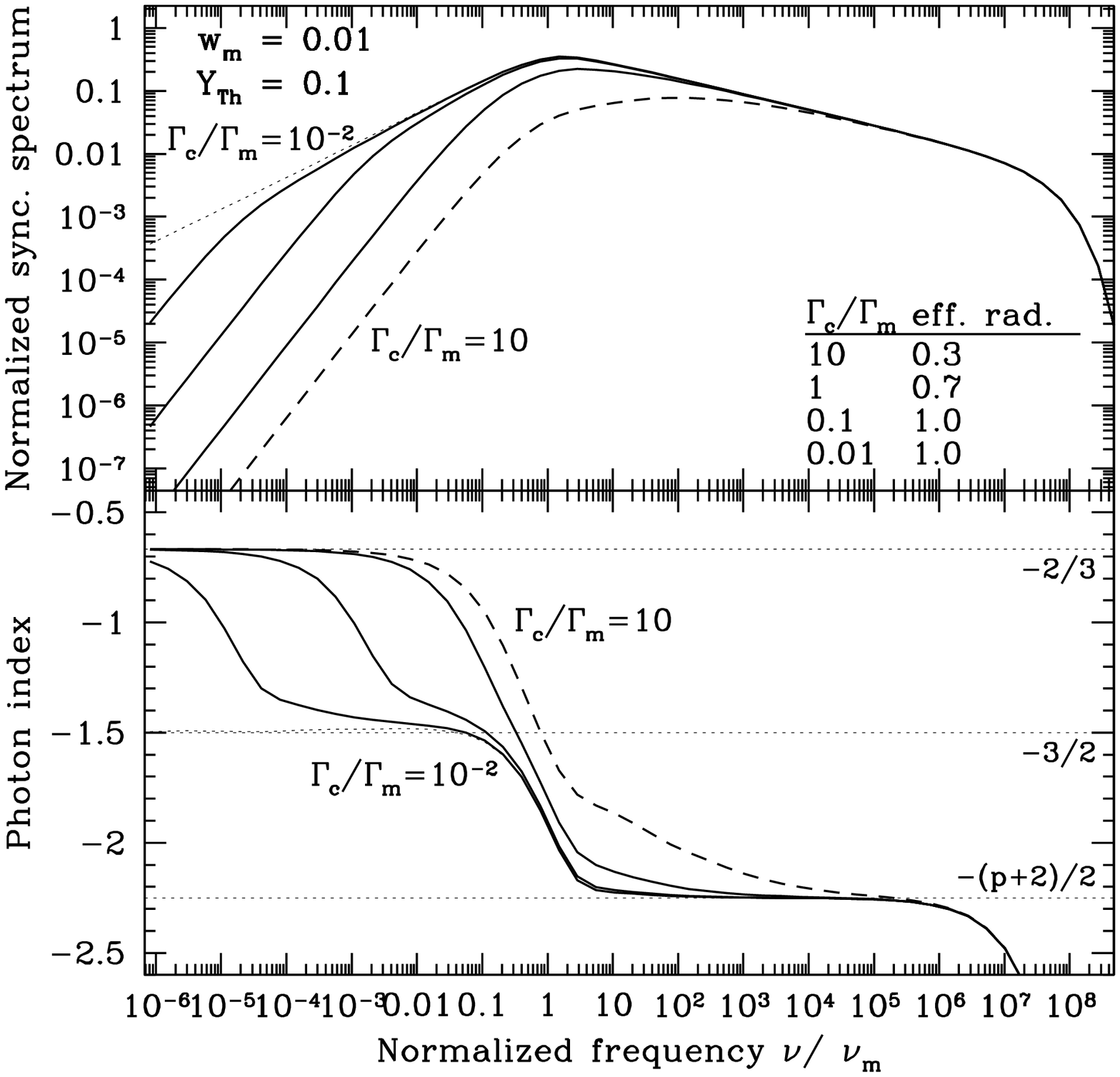} & \includegraphics[width=0.47\textwidth]{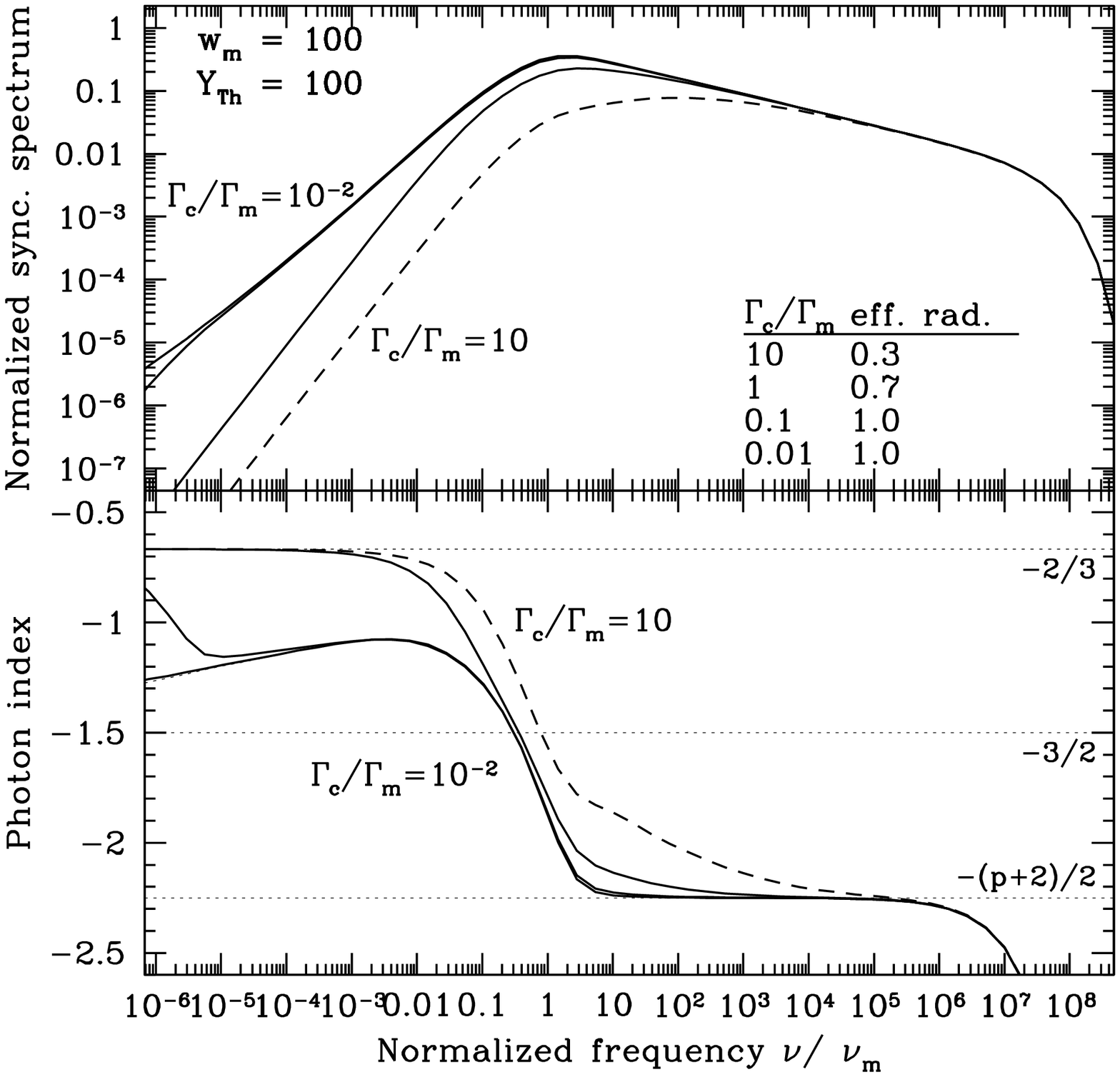} \\
\includegraphics[width=0.47\textwidth]{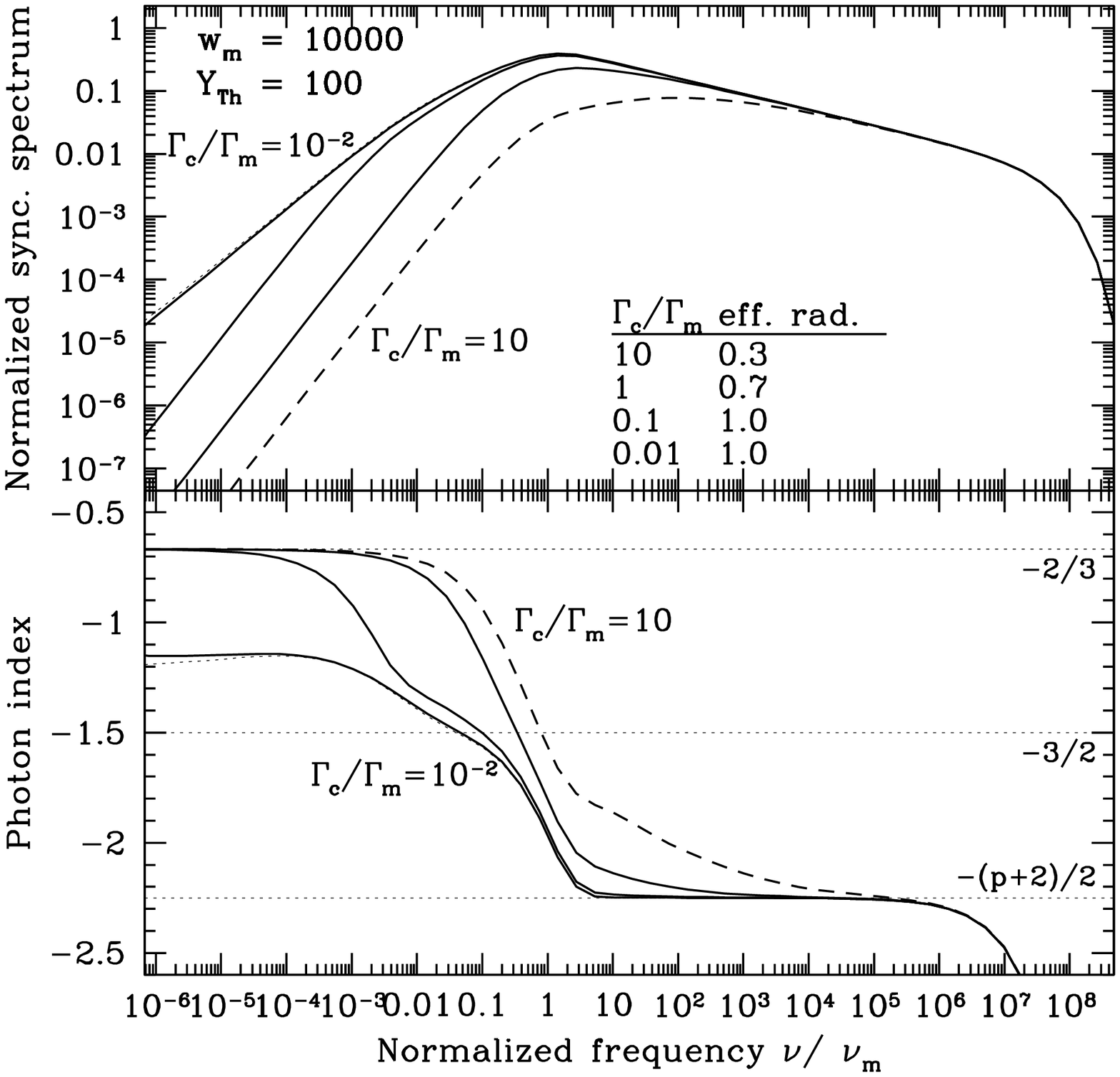} & \includegraphics[width=0.47\textwidth]{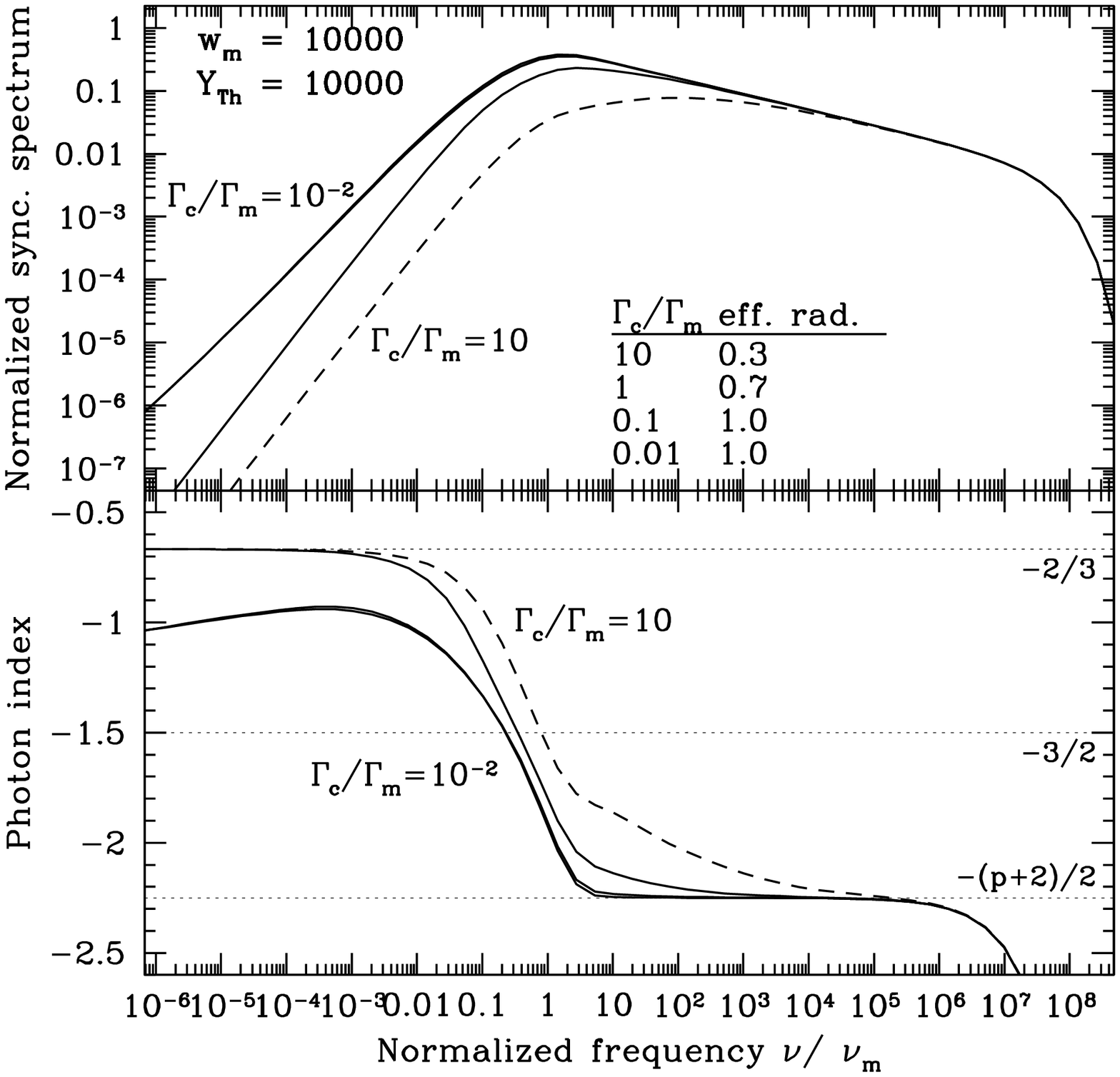} 
\end{tabular}
\end{center}
\caption{\textbf{The effect of adiabatic cooling on the fast cooling synchrotron spectrum in presence of inverse Compton scatterings.} 
The normalized synchrotron spectrum defined by $\nu \left.u_{\nu}\right|_\mathrm{syn}/ n_\mathrm{e}^\mathrm{acc}\Gamma_\mathrm{m}m_\mathrm{e}c^2$ is plotted as a function of the normalized frequency $\nu/\nu_\mathrm{m}$, as well as the corresponding photon index $d\ln{u_{\nu}}/d\ln{\nu}-1$.
All spectra in thick solid line are computed numerically using a detailed radiative code including synchrotron radiation, inverse Compton scatterings and adiabatic cooling (see text). Spectra in thin dotted line are computed without inverse Compton scatterings. All other processes (synchrotron self-absorption, $\gamma\gamma$ annihilation) are neglected. Each panel corresponds to a different set of parameters $\left(w_\mathrm{m},Y_\mathrm{Th}\right)$ indicated in the top-left corner. In each panel, spectra are plotted for increasing ratios $\Gamma_\mathrm{c}/\Gamma_\mathrm{m}=0.01$, $0.1$ , $1$ and $10$. As in \reffig{fig:effectKN}, the maximum Lorentz factor of electrons is computed with a fixed ratio $\Gamma_\mathrm{max}/\Gamma_\mathrm{min}=10^4$. Spectra in dashed lines are radiatively inefficient (slow cooling regime). The table inserted in each panel lists the values of the radiative efficiency $f_\mathrm{rad}$ of the electrons. 
}
\label{fig:effectAC}
\end{figure*}

 To illustrate this situation we plot in \reffig{fig:effectAC} the evolution of the synchrotron spectrum for an increasing ratio $\Gamma_\mathrm{c}/\Gamma_\mathrm{m}$ and for different values of $\left(w_\mathrm{m};Y_\mathrm{Th}\right)$ representative of the different regions in the diagram of \reffig{fig:diagKN}. As expected a break in the spectrum appears at frequency $\nu_\mathrm{c,eff}$, i.e. at the synchrotron frequency of electrons with Lorentz factor $\Gamma_\mathrm{c,eff}$, whose radiative timescale is equal to the dynamical timescale $t_\mathrm{ex}$. We have $\Gamma_\mathrm{c,eff}\le \Gamma_\mathrm{c}$ due to inverse Compton scatterings.The photon index below $\nu_\mathrm{c,eff}$ is $-2/3$. Therefore, when $\nu_\mathrm{c,eff}$ is close to $\nu_\mathrm{m}$ the observed photon index can be very close to this asymptotic value, even in fast cooling regime. This is well seen in \reffig{fig:effectAC} for $w_\mathrm{m}=100$ and $Y_\mathrm{Th}=100$ (top-right panel) or $w_\mathrm{m}=10^4$ and $Y_\mathrm{Th}=10^4$ (bottom-right panel) and for $\Gamma_\mathrm{c}/\Gamma_\mathrm{m}=1$ ($f_\mathrm{rad}\simeq 0.6-0.7$ in this case). When the efficiency of inverse Compton scatterings is reduced,  $\nu_\mathrm{c,eff}$ is closer to $\nu_\mathrm{c}$ and the same effect can be seen for lower values of the ratio $\Gamma_\mathrm{c}/\Gamma_\mathrm{m}$. This is the case for instance for $w_\mathrm{m}=0.01$ and $Y_\mathrm{Th}=0.1$ (top-left panel) or for $w_\mathrm{m}=10^4$ and $Y_\mathrm{Th}=100$ (bottom-left panel) and for $\Gamma_\mathrm{c}/\Gamma_\mathrm{m}=0.1-1$ ($f_\mathrm{rad}\simeq 0.6-0.7\,\to 1$ in this case).
 
\begin{figure*}[t]
\begin{center}
\begin{tabular}{cc}
\includegraphics[width=0.47\textwidth]{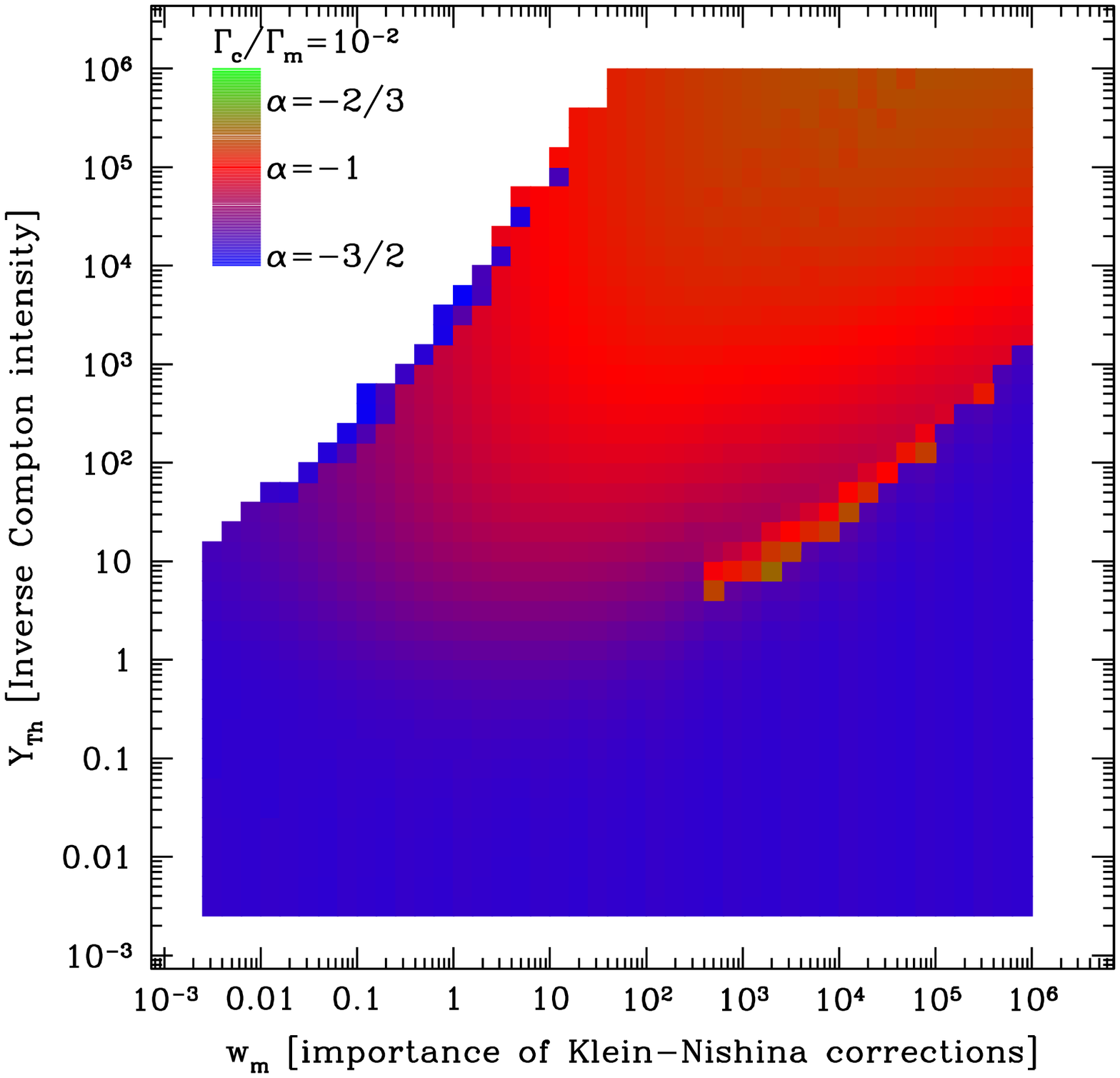} & \includegraphics[width=0.47\textwidth]{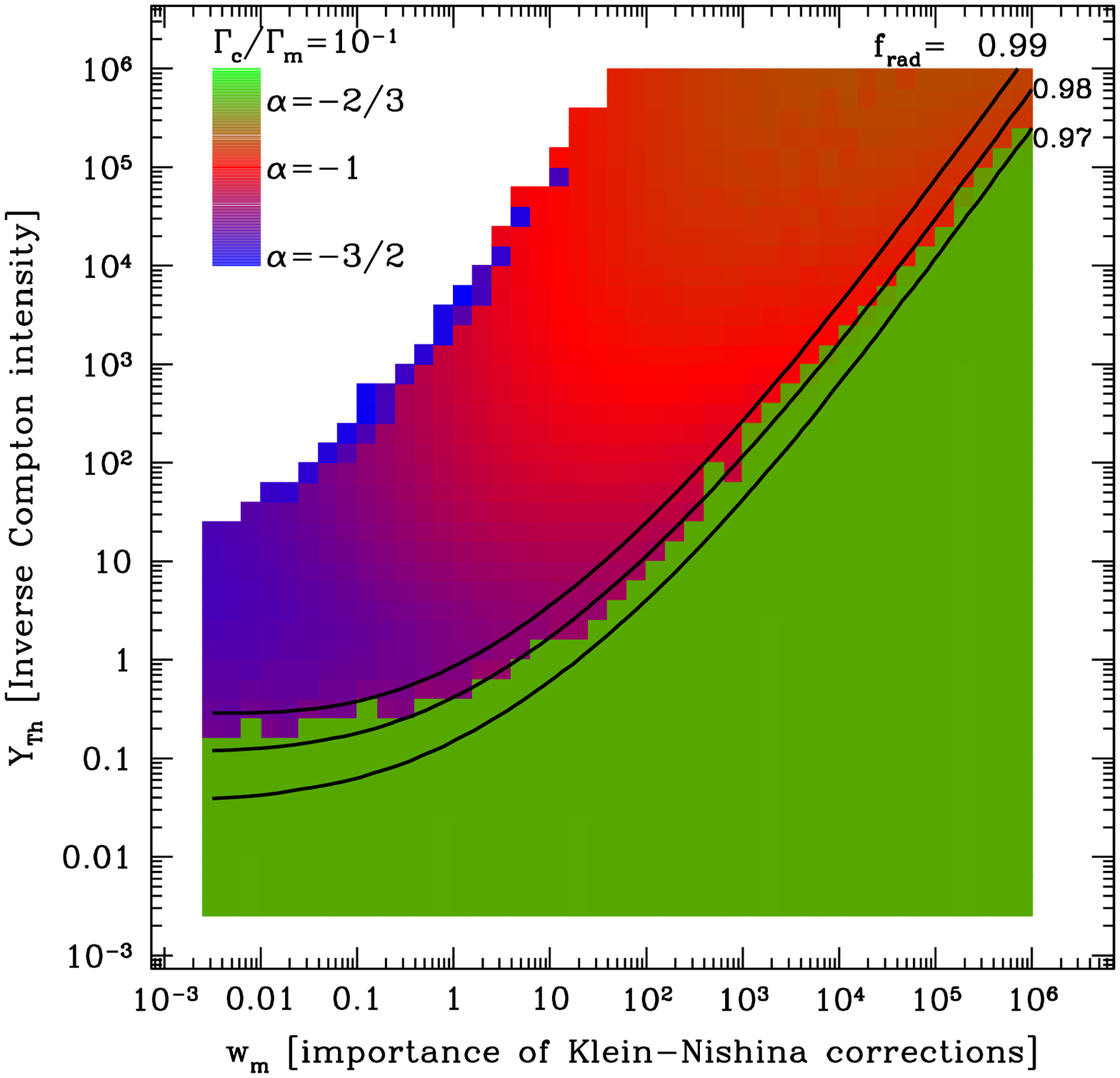}\\
\includegraphics[width=0.47\textwidth]{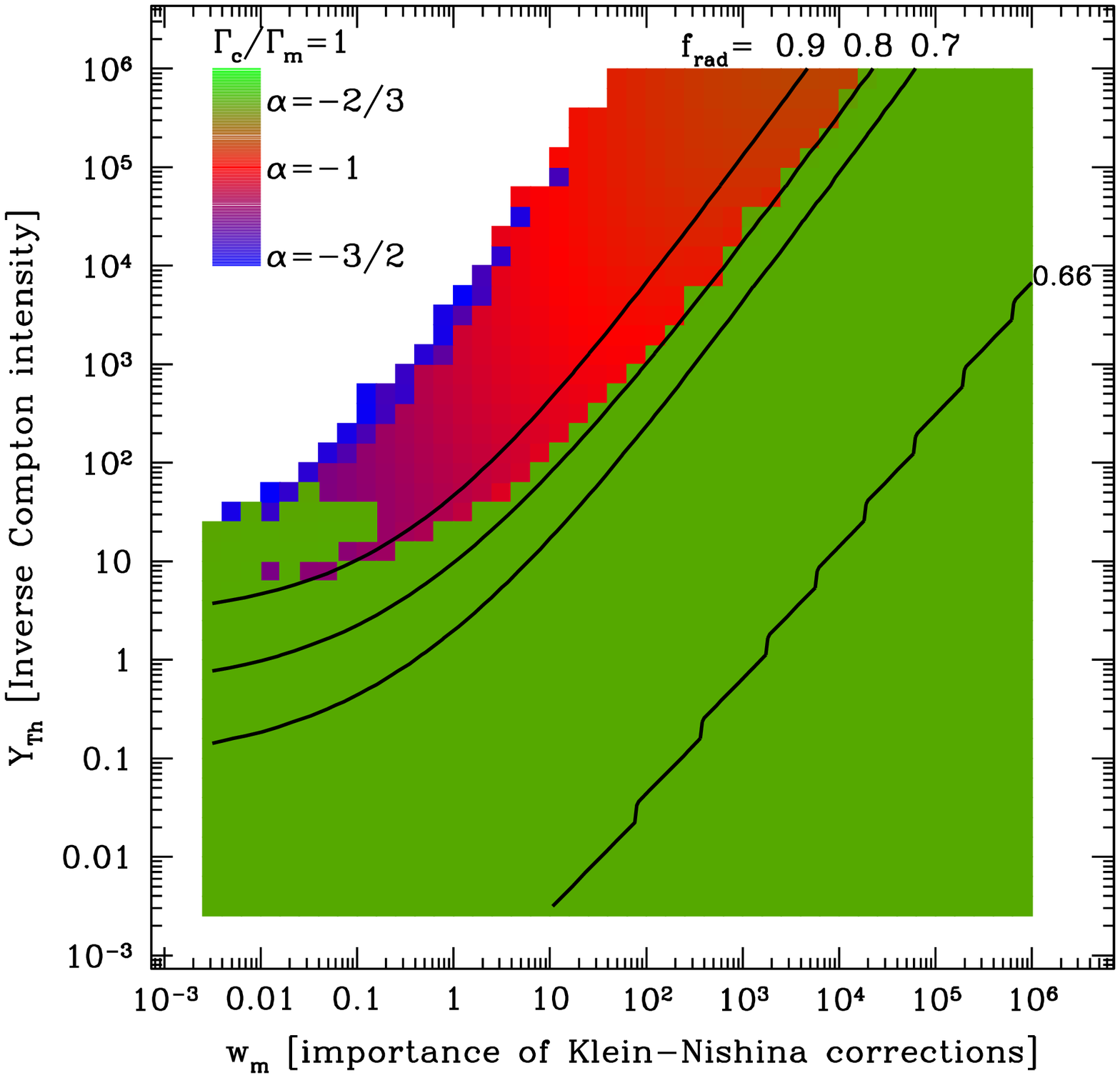} & \includegraphics[width=0.47\textwidth]{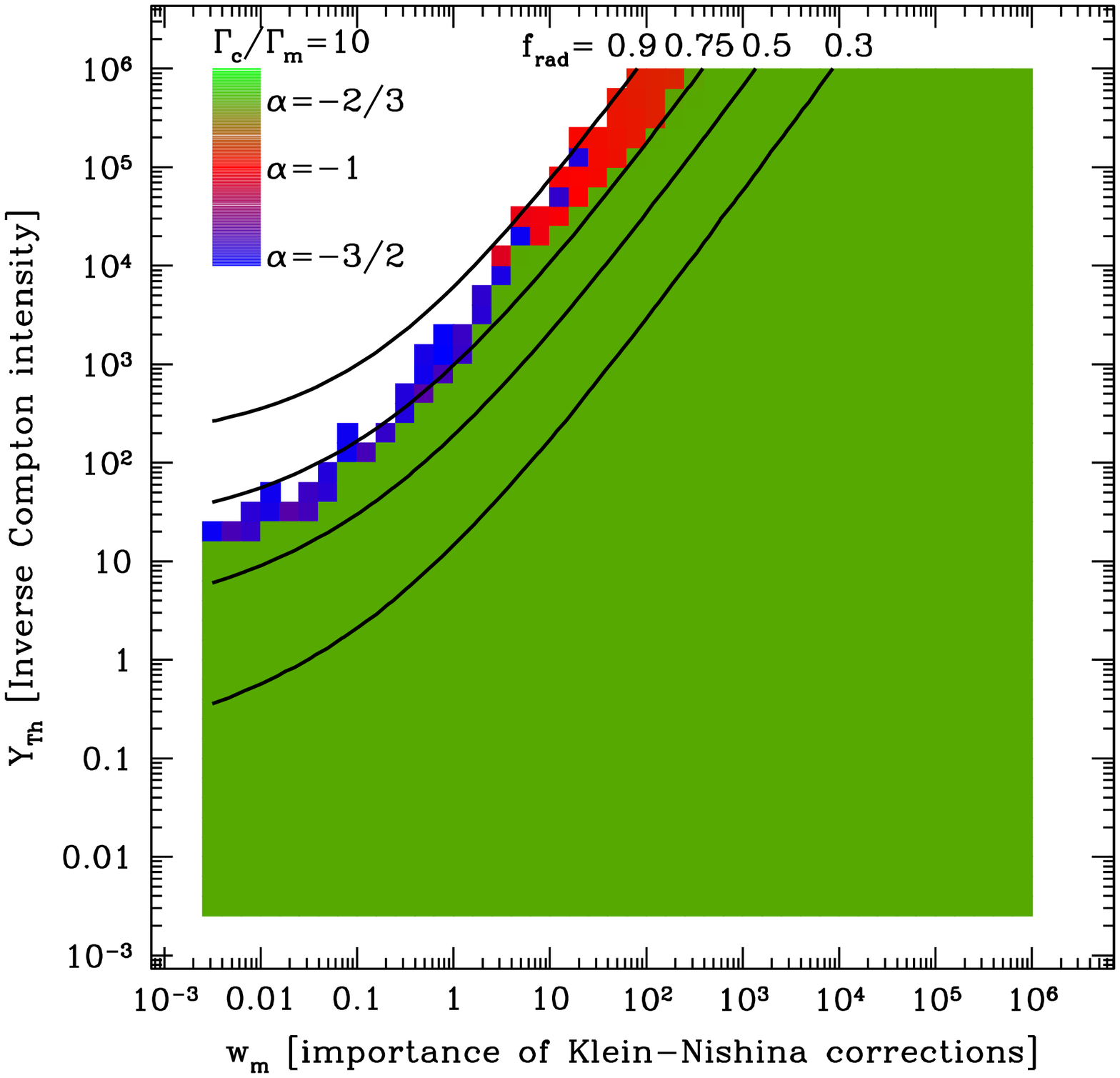}\\
\end{tabular}
\end{center}
\caption{\textbf{The low-energy slope $\alpha$ of the synchrotron spectrum in the presence of inverse Compton scatterings in Klein-Nishina regime, including the effect of adiabatic cooling.} Same as \reffig{fig:diagKN}, now including adiabatic cooling for $\Gamma_\mathrm{c}/\Gamma_\mathrm{m}=10^{-2}$, $10^{-1}$, $1$ and $10$. Black solid lines of constant radiative efficiency $f_\mathrm{rad}$ are plotted on top of the three last diagrams. In the first panel, the radiative efficiency is always close to $100\,\%$. For larger ratio $\Gamma_\mathrm{c}/\Gamma_\mathrm{m}$, the radiative efficiency remains larger than respectively $96\,\%$ ($\Gamma_\mathrm{c}/\Gamma_\mathrm{m}=10^{-1}$), $65\,\%$   ($\Gamma_\mathrm{c}/\Gamma_\mathrm{m}=1$) and $28\,\%$ ($\Gamma_\mathrm{c}/\Gamma_\mathrm{m}=10$). }
\label{fig:diagAC}
\end{figure*}

We have plotted in \reffig{fig:diagAC} the same diagram as in \reffig{fig:diagKN}, now including the effect of adiabatic cooling for different values of the ratio $\Gamma_\mathrm{c}/\Gamma_\mathrm{m}$. The representative value of $\alpha$ is selected in the same way as in \reffig{fig:diagKN}, but the maximum is not always as clearly defined as in the $\Gamma_\mathrm{c}/\Gamma_\mathrm{m}=0$ case (see for instance the curves for $\Gamma_\mathrm{c}/\Gamma_\mathrm{m}=0.1$ in the four panels of \reffig{fig:effectAC}). Clearly, when the intermediate maximum of the spectral slope below $\nu_\mathrm{m}$ disappears and $\alpha$ is given the asymptotic value $-2/3$ (green region in \reffig{fig:diagAC}), the comparison with observations of the low-energy photon index becomes difficult :
depending on the location of the peak energy $E_\mathrm{p}$, and of the low-energy threshold of the instrument, any value of $\alpha$ between $-2$ and $-2/3$ can be measured.\\
 The  diagrams    
in \reffig{fig:diagAC} show that already
for $\Gamma_\mathrm{c}/\Gamma_\mathrm{m}=0.1$, the asymptotic slope $\alpha=-2/3$ is reached in a large region of the $w_\mathrm{m}-Y_\mathrm{Th}$ plane, 
together with a large radiative efficiency. Even for $\Gamma_\mathrm{c}/\Gamma_\mathrm{m}=1$, $\alpha=-2/3$ is observed in a large region where the radiative efficiency is still larger than $66\%$. It is only for $\Gamma_\mathrm{c}/\Gamma_\mathrm{m}=10$ that the slow cooling regime dominates the diagram, high radiative efficiency being found together with $\alpha=-2/3$ in only a very small area.\\

%%%%%%%%%%%%%%%%%%%%%%%%%%%%%%%%%%%%%%%%%%%%%%%%%%%%%%%%%
% Figure 4
%
%%%%%%%%%%%%%%%%%%%%%%%%%%%%%%%%%%%%%%%%%%%%%%%%%%%%%%%%%

%%%%%%%%%%%%%%%%%%%%%%%%%%%%%%%%%%%%%%%%%%%%%%%%%%%%%%%%%
% Figure 5
%
%%%%%%%%%%%%%%%%%%%%%%%%%%%%%%%%%%%%%%%%%%%%%%%%%%%%%%%%%

\subsubsection{Other effects}
Other processes may influence the final spectral shape. Photon-photon annihilation produces a cutoff at high energy \citep{granot:08,bosnjak:09}. For all cases presented in this paper, we have checked that the opacity for this process was extremely low below $\sim 100\, \mathrm{MeV}$. Photon photon annihilation can affect the tail of the inverse Compton component at high-energies. The fraction of the radiated energy which is reinjected in pairs via $\gamma\gamma\to e^+e^-$ is usually, but not always, small in the examples shown in \refsec{sec:internalshocks}. For example, in the three cases defined in \refsec{sec:internalshocks}, it is typically less than $10^{-3}$ in case C, less than $0.05$ in case A and between $0.1$ and $0.3$ in case B. For collisions occuring at low radius or with low Lorentz factors, $\gamma\gamma$ annihilation could be even more important. When the fraction of the energy in annihilated photons is non negligible, the resulting radiation of the created pairs could affect the spectral shape even at low energy, and modify the results presented here. Despite its potentially interesting impact, we defer to future work the investigation of such cases, due to the additional complexity it involves for the computation of the radiated spectrum. Numerical approaches to solve such a highly non linear problem including thermalization effects have been proposed by \citet{peer:05,asano:07,belmont:08,vurm:09}\\

At low energy, the synchrotron self-absorption can also steepen the spectrum. This effect is included in simulations presented in \refsec{sec:internalshocks} and is always negligible in the soft gamma-ray domain, in agreement with the standard predictions for the synchrotron fast cooling regime. Indeed the timescale for self-absorption at $\nu_\mathrm{m}$ is given by \citep[see e.g. equation (28) in][]{bosnjak:09} :
\begin{equation}
\frac{t_\mathrm{a}\left(\nu_\mathrm{m}\right)}{t_\mathrm{ex}} \simeq \frac{4\pi \nu_\mathrm{m}^3}{n_\mathrm{e}^\mathrm{acc} c^3} \simeq
1.3\times 10^{14}\, \left(\frac{\nu_\mathrm{m}}{1\, \mathrm{keV}}\right)^3\left(\frac{t_\mathrm{ex}}{1\, \mathrm{s}}\right)\left(\frac{\tau_\mathrm{e}^\mathrm{acc}}{10^{-6}}\right)^{-1}\, ,
\end{equation}
where $1\, \mathrm{keV}$ is taken for a typical value of the peak energy in the comoving frame and other parameters are given representative values for internal shocks. The timescale for self-absorption around $\nu_\mathrm{m}$ is therefore always much larger than all other timescales (dynamical or radiative) and the self-absorption process is negligible in the soft gamma-ray range.

In addition to the details of the radiative processes, the precise shape of the electron distribution can also have an impact on the final spectrum. Here, we assume a power-law distribution. More complex distributions showing several components (e.g. Maxwellian distribution + non-thermal tail) are observed in some simulations of particle acceleration in ultra-relativistic shocks \citep{spitkovsky:08a,spitkovsky:08b,martins:09}. Such results would need to be confirmed for the mildly relativistic regime of interest for the prompt GRB emission. In the ultra-relativistic regime relevant for the afterglow, \citet{giannios:09} have shown that the Maxwellian component could have an observable signature. However \citet{baring:04} find that the non-thermal electron population should dominate in the prompt phase. 
We leave to a future work the study of the consequences of more complex electron distributions on the observed GRB prompt spectra.

% =======================================================
% SECTION 4 : Discussion in the framework of the internal shock model
% =======================================================

\section{Constraints on the internal shock model}
\label{sec:internalshocks}

% SECTION 4.1 : Exploration of the parameter space of internal shocks
% ---------------------------------------------------------------------------------------

\subsection{General constraints on the physical conditions in the emitting regions}
\label{sec:GRBparameters}
We have shown in \refsec{sec:theory} that the spectrum resulting from synchrotron radiation in the presence of inverse Compton scatterings in Klein-Nishina regime can account for observed low-energy photon index $\alpha=-3/2$ to $-1$, and that the additional effect of adiabatic cooling with $\Gamma_\mathrm{c}\la\Gamma_\mathrm{m}$ can lead to steeper slopes up to $\alpha=-2/3$ in a regime where the radiative efficiency is still reasonably high ($f_\mathrm{rad}\ga 50\,\%$). In principle, this allows to reconcile the synchrotron process with the observed spectral parameters in most GRB spectra (see \refsec{sec:catalog}). However it is still necessary to demonstrate that the physical conditions identified in \refsec{sec:theory} can be reached in the emitting regions within GRB outflows. These conditions are approximatively  $w_\mathrm{m}\ga 1$ and $w_\mathrm{m}^{1/3}\le Y_\mathrm{Th}\la w_\mathrm{m}^{3}$ for changing the slope from $\alpha=-3/2$ to $-1$ with the effect of IC scatterings in Klein-Nishina regime, and $\Gamma_\mathrm{m}/\Gamma_\mathrm{c}\simeq 0.1-1$ to get steeper slopes up to $\alpha=-2/3$ with the effect of adiabatic cooling.\\ 

%%%%%%%%%%%%%%%%%%%%%%%%%%%%%%%%%%%%%%%%%%%%%%%%%%%%%%%%%
% Figure 6
%
\begin{figure}
\begin{center}
\includegraphics[width=0.5\textwidth]{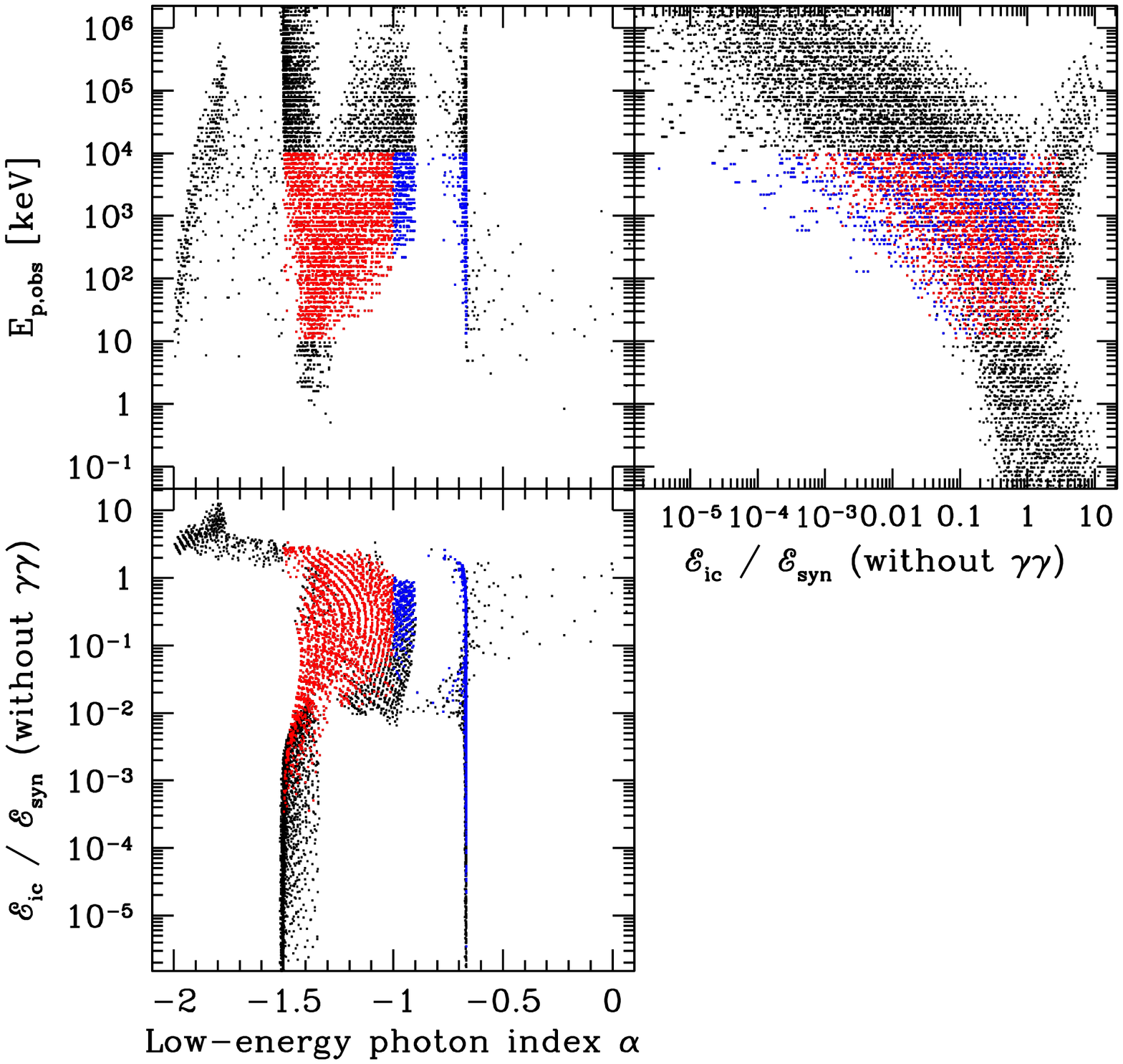}\\
\end{center}
\caption{\textbf{Exploration of the internal shock parameter space (1).} 
All cases fulfilling the conditions of (i) transparency; (ii) radiative efficiency; and (iii) synchrotron dominance at low-energy (see text) are plotted with black dots in three planes : $E_\mathrm{p,obs}$ vs $\alpha$ (top-left panel); $\mathcal{E}_\mathrm{ic}/\mathcal{E}_\mathrm{syn}$ vs $\alpha$ (bottom-left panel); $E_\mathrm{p,obs}$ vs $\mathcal{E}_\mathrm{ic}/\mathcal{E}_\mathrm{syn}$ (top-right panel). The ratio $\mathcal{E}_\mathrm{ic}/\mathcal{E}_\mathrm{syn}$ does not take into account the fraction of high energy photons that are suppressed by $\gamma\gamma$ annihilation. In addition, the cases where the synchrotron spectrum peaks in the gamma-ray range  ($10\,\mathrm{keV}\le E_\mathrm{p,obs}\le 10\,\mathrm{MeV}$) and the low-energy photon index is in the range $-3/2\le\alpha\le-1$ (resp. $-1\le\alpha\le-2/3$)
are plotted in red (resp. blue).
}
\label{fig:explo1}
\end{figure}
%%%%%%%%%%%%%%%%%%%%%%%%%%%%%%%%%%%%%%%%%%%%%%%%%%%%%%%%%

%%%%%%%%%%%%%%%%%%%%%%%%%%%%%%%%%%%%%%%%%%%%%%%%%%%%%%%%%
% Figure 7
%
\begin{figure*}
\begin{center}
\begin{tabular}{ccc}
\includegraphics[width=0.3\textwidth]{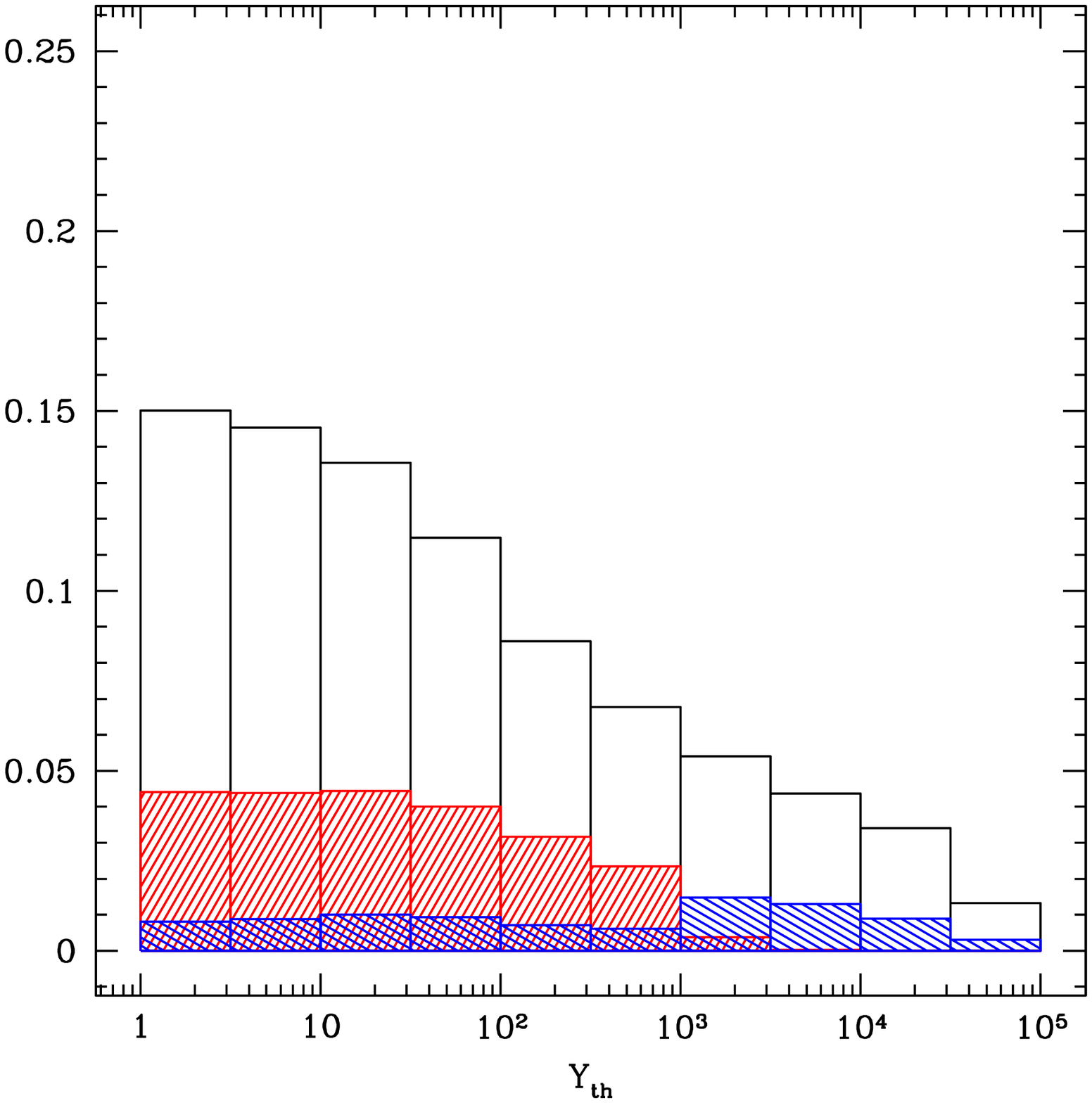} &
\includegraphics[width=0.3\textwidth]{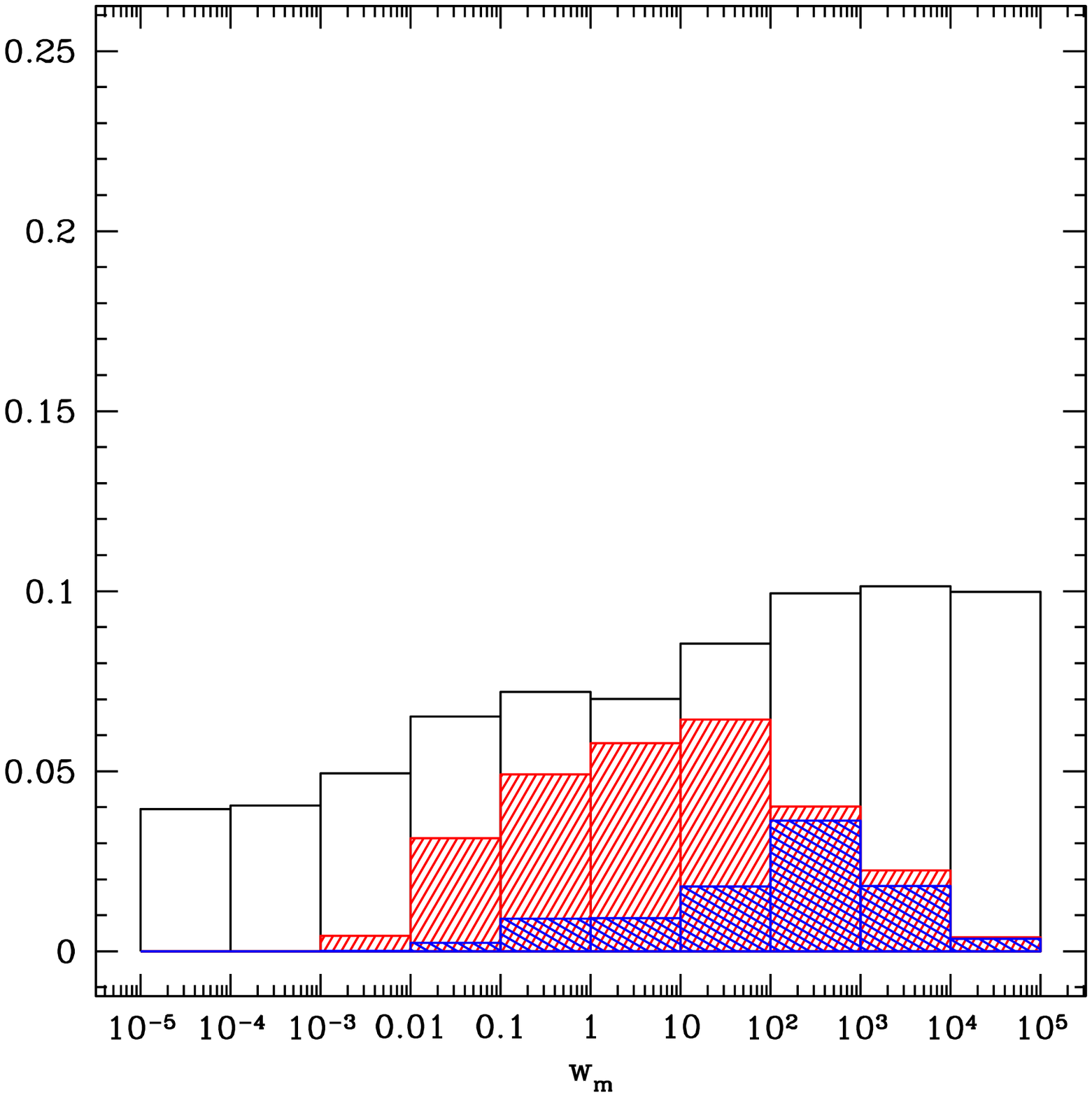} &
\includegraphics[width=0.3\textwidth]{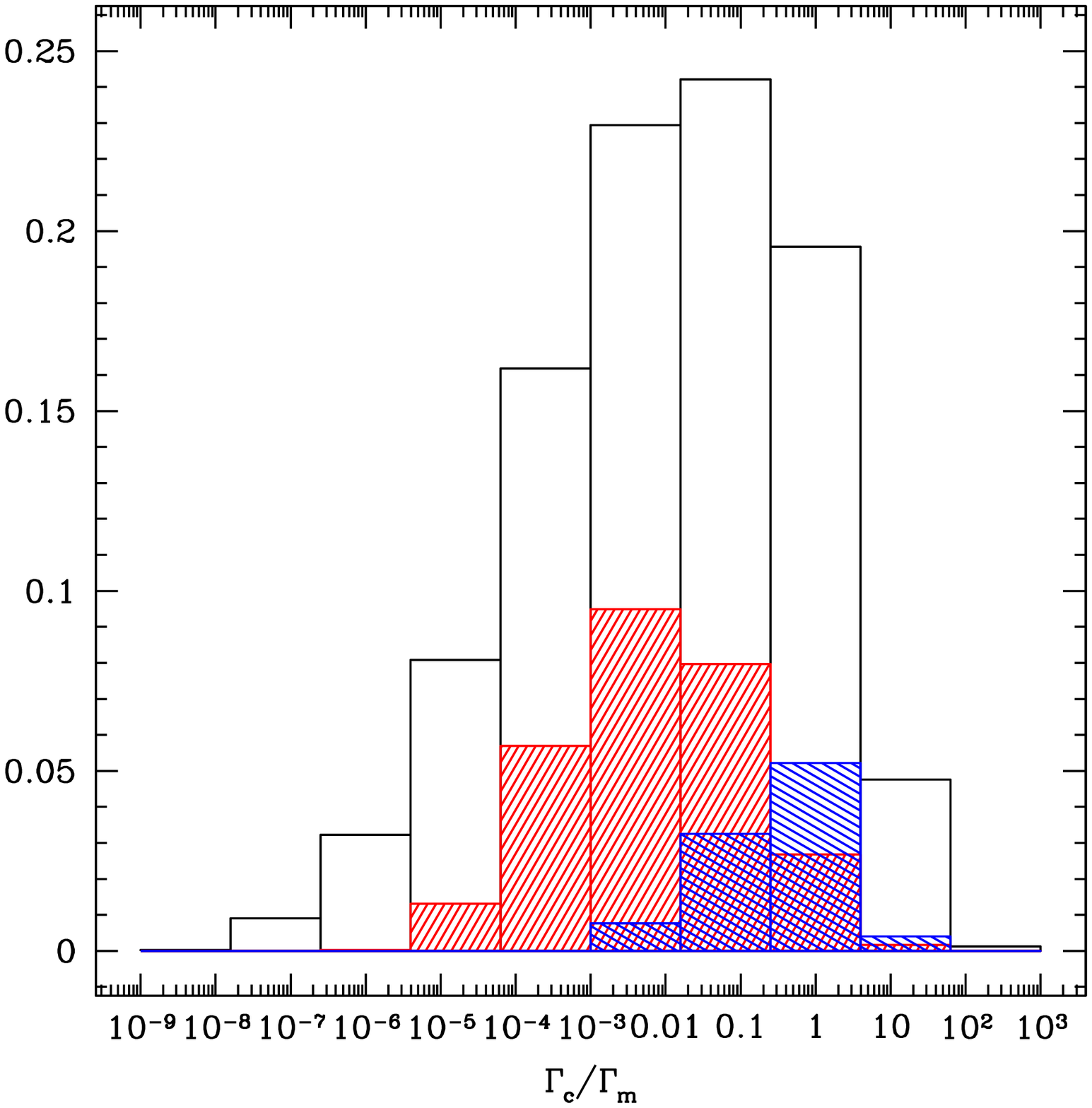}\\
\end{tabular}
\end{center}
\caption{\textbf{Exploration of the internal shock parameter space (2).} 
The distribution of the three parameters defining the relative importance of inverse Compton scatterings, Klein-Nishina corrections and adiabatic cooling, $Y_\mathrm{th}$, $w_\mathrm{m}$ and $\Gamma_\mathrm{c}/\Gamma_\mathrm{m}$ are plotted for all cases
presented in \reffig{fig:explo1}. In addition the distributions of the same parameters for cases where the synchrotron spectrum peaks in the gamma-ray range ($10\,\mathrm{keV}\le E_\mathrm{p,obs}\le 10\,\mathrm{MeV}$) and the low-energy photon index is in the correct range are also plotted ($-3/2\le\alpha\le -1$: red; $-1\le\alpha\le-2/3$: blue).  
}
\label{fig:explo2}
\end{figure*}
%%%%%%%%%%%%%%%%%%%%%%%%%%%%%%%%%%%%%%%%%%%%%%%%%%%%%%%%%

We assume that the prompt gamma-ray emission is produced in a relativistic outflow ejected by a source at redshift $z$. We consider an emitting region at radius $R$ within the outflow, with a Lorentz factor $\Gamma_*$. We do not specify at this stage the physical mechanism responsible for the energy dissipation in this region, leading to the presence of a magnetic field $B$ and a population of relativistic electrons with a minimum Lorentz factor $\Gamma_\mathrm{m}$. We assume that the medium is optically thin, i.e. $\tau_\mathrm{T}=\sigma_\mathrm{T} n_\mathrm{e}^\mathrm{acc} c t_\mathrm{ex}\le 1$.
The observed peak of the synchrotron spectrum is given by
\begin{equation}
(1+z)h\nu_\mathrm{m,obs} \simeq 380\,\mathrm{keV}\,\left(\frac{\Gamma_{*}}{300}\right)\left(\frac{\Gamma_\mathrm{m}}{10^4}\right)^2\left(\frac{B}{3000\,\mathrm{G}}\right)\ .
\label{eq:numobs}
\end{equation}
The dynamical timescale relevant for adiabatic cooling can be estimated by
\begin{equation}
t_\mathrm{ex} = 1.1\,\mathrm{s}\,\left(\frac{R}{10^{13}\ \mathrm{cm}}\right)\left(\frac{\Gamma_*}{300}\right)^{-1}\ .
\end{equation}
From \refeq{eq:gc}, this leads to
\begin{equation}
\frac{\Gamma_\mathrm{c}}{\Gamma_\mathrm{m}} \simeq 0.78\,
\left(\frac{R}{10^{13}\ \mathrm{cm}}\right)^{-1}\left(\frac{\Gamma_*}{300}\right)\left(\frac{B}{3000\ \mathrm{G}}\right)^{-2}\left(\frac{\Gamma_\mathrm{m}}{10^4}\right)^{-1}\ .
\label{eq:estimgc}
\end{equation}
The two parameters governing inverse Compton scatterings are given by \refeq{eq:wm} and \refeq{eq:yth}\,:
\begin{equation}
w_\mathrm{m} \simeq 100 \left(\frac{\Gamma_\mathrm{m}}{10^4}\right)^3\left(\frac{B}{3000\,\mathrm{G}}\right)
\label{eq:estimwm}
\end{equation}
and
\begin{equation}
Y_\mathrm{Th} \simeq 100 \left(\frac{\tau_\mathrm{T}}{10^{-6}}\right)
\left(\frac{R}{10^{13}\ \mathrm{cm}}\right)^{-1}\left(\frac{\Gamma_*}{300}\right)\left(\frac{B}{3000\ \mathrm{G}}\right)^{-2}\left(\frac{\Gamma_\mathrm{m}}{10^4}\right)\ .
\label{eq:estimyth}
\end{equation}
Lorentz factors $\Gamma_*$ above $100$ are necessary to avoid the presence of a high-energy cutoff in the spectrum due to $\gamma\gamma$ annihimation \citep[see e.g.][]{lithwick:01}. Higher values are even required in some bursts detected by \textit{Fermi}/LAT. For instance $\Gamma\ga 600-900$ has been derived by the \textit{Fermi}/LAT collaboration for GRB 080916C \citep{abdo:09}. Radii in the range $10^{13}$--$10^{15}\,\mathrm{cm}$ are expected as the prompt GRB emission should mainly occur above the photospheric radius and below the deceleration radius. Then, the main constraint comes from the fact that in the proposed scenario gamma-ray photons must be produced directly by synchrotron emission. From \refeq{eq:numobs}, this is always possible if electrons can be accelerated to very high Lorentz factors. 
Then \refeq{eq:estimwm} and (\ref{eq:estimyth}) show that 
the physical conditions listed above and leading to $-3/2\le\alpha\le-1$ can indeed be reached in GRBs. In addition \refeq{eq:estimgc}
indicates that the "marginally fast cooling regime" leading to $\alpha\to -2/3$ can in principle also be reached, especially for emission at larger radius, where the magnetic field could be lower.  

% SECTION 4.2 : Exploration of the parameter space of internal shocks
% ---------------------------------------------------------------------------------------

\subsection{Impact on the microphysics parameters in internal shocks}
The internal shock model allows a self-consistent evaluation of the physical conditions (i.e. $\Gamma_*$, $R$, $B$, $\Gamma_\mathrm{m}$, etc.) for each emitting region. 
It is then possible
to identify the pertinent range of the parameters 
leading to steep low-energy slopes : the properties of the relativistic outflow (Lorentz factor, kinetic energy) and the parameters describing the microphysics at work in shocked regions (particle acceleration, magnetic field amplification).\\

We consider first collisions between two equal-mass shells \citep{barraud:05,bosnjak:09}. More realistic outflows are considered in the next subsection.  The parameters defining the dynamical properties of a collision are the mean Lorentz factor in the outflow $\bar{\Gamma}$, the ratio $\kappa$ of the Lorentz factor of the faster shell over the Lorentz factor of the slower shell, the time separation $\tau$ between the ejection of the two shells and the kinetic energy flux $\dot{E}$ injected in the outflow. These 4 parameters allow to estimate the radius of the collision, as well as the physical conditions in the shocked region (Lorentz factor $\Gamma_*$, comoving density $\rho_*$ and comoving specific energy density $\epsilon_*$). In addition, four microphysics parameters are necessary to estimate the distribution of relativistic electrons and the magnetic field : $\epsilon_\mathrm{B}$ and $\epsilon_\mathrm{e}$ are the fraction of the energy density $\rho_*\epsilon_*$ that is injected in the magnetic field and the relativistic electrons, respectively. We assume that only a fraction $\zeta$ of the available electrons are accelerated in a non-thermal distribution and that this distribution is a power-law with index $-p$. The full description of this  model can be found in \citet{bosnjak:09}.\\

We explore the parameter space of this model, assuming a constant value $\epsilon_\mathrm{e}=1/3$ and a constant electron slope $p=2.5$. A high value of $\epsilon_\mathrm{e}$ seems unavoidable in internal shocks to maintain a reasonable efficiency of the process. We have checked that our results are not  affected much by taking $\epsilon_\mathrm{e}=0.1$. As we are mainly interested in the low-energy photon index $\alpha$, the choice of $p$ is not very important. The present value $p=2.5$, leads to a high-energy photon index $\beta\simeq -2.25$ close to the mean value observed in BATSE spectroscopic catalog \citep{preece:00}. We then compute the observed spectrum of 50400 collisions for $\log{\bar{\Gamma}}=$1.5, 2, 2.5 and 3; $\kappa=$2.5, 5, 7.5 and 10; $\log{\tau}=$-2, -1, 0, 1 and 2; $\log{\dot{E}}=$50, 51, 52, 53, 54 and 55; $\log{\zeta}=$-4, -3, -2, -1 and 0; $\log{\epsilon_\mathrm{B}}=-5.5\to -0.5$ every 0.25. These spectra are computed with the radiative code developed by \citet{bosnjak:09}, including all relevant processes : synchrotron radiation, inverse Compton scatterings and adiabatic cooling, as considered in the previous section, but also synchrotron self-absorption and $\gamma\gamma$ annihilation.\\

We keep only cases which fulfill the following conditions\,: (i) the shocked region is transparent ($\sigma_\mathrm{T} n_\mathrm{\pm} c t_\mathrm{ex} < 0.1$, where $n_{\pm}$ is the final lepton density in the shocked region, taking into account pairs that were created by $\gamma\gamma$ annihilation but neglecting pair annihilation, see \citealt{bosnjak:09}) ; (ii) electrons are radiatively efficient ($f_\mathrm{rad}>0.5$) ; (iii) synchrotron radiation is dominant at low-energy ($\left.u_\mathrm{\nu}\right|_\mathrm{syn} > 10 \left. u_\mathrm{\nu} \right|_\mathrm{ic}$ at the frequency of the synchrotron peak). This last condition eliminates a few cases where inverse Compton scatterings are so efficient that the synchrotron component is hardly observed. After this selection, $\sim 75\%$
of cases are suppressed, because of a too large optical depth (condition (i): $\sim 50\%$ of cases), 
 a too low radiative efficiency (condition (ii): $\sim 20 \%$ of cases) or a negligible synchrotron emission (condition (iii): $\sim 10 \%$ of cases). 
For each spectrum, assuming a source redshift $z=1$, we compute the observed peak energy of the synchrotron component $E_\mathrm{p,obs}$, the low-energy photon index $\alpha$ below the peak, and the ratio $\mathcal{E}_\mathrm{ic}/\mathcal{E}_\mathrm{syn}$ of the inverse Compton component over the synchrotron component. All models fulfilling the three conditions listed above are plotted in \reffig{fig:explo1}. 
This figure illustrates that the internal shock model with a dominant synchrotron process in the soft gamma-ray range (BATSE, \textit{Fermi}/GBM) allows a large range of low-energy photon index between $-3/2$ and $-2/3$ and a large range of peak energies (including very high peak energies as observed in some bright bursts such as GRB 080916C, \citealt{abdo:09}, for which \citealt{wang:09} have recently shown that the slope $\alpha\simeq -1$ can be explained by the effect of IC scatterings in KN regime on the synchrotron spectrum; and very low peak energies as expected in X-ray rich gamma-ray bursts or X-ray flashes, \citealt{heise:01,sakamoto:05,sakamoto:08}). All these cases have $f_\mathrm{rad}>0.5$ and even $f_\mathrm{rad}>0.9$ in half  of the cases.
For cases with $10\,\mathrm{keV}\le E_\mathrm{p,obs}\le10\,\mathrm{MeV}$, the ratio $\mathcal{E}_\mathrm{ic}/\mathcal{E}_\mathrm{syn}$  is typically in the range $10^{-3}$--$1$. This is in agreement with the indication from the \textit{Fermi}/LAT GRB detection rate that most GRBs do not have a strong additional component between 100 MeV and 10 GeV \citep{granot:10a}. Note that the density of points in \reffig{fig:explo1} has no physical meaning as the distribution of the physical parameters in GRB outflows is unknown. This figure only illustrates the range of observed values that can be expected in the internal shock model.\\

We plot in \reffig{fig:explo2} the distributions of $w_\mathrm{m}$, $Y_\mathrm{Th}$ and $\Gamma_\mathrm{c}/\Gamma_\mathrm{m}$ for the same models. The values of the low-energy photon index $\alpha$ are in full agreement with the analysis made in section~\refsec{sec:theory}. How are such values obtained ? The distributions of dynamical ($\bar{\Gamma}$, $\kappa$, $\tau$ and $\dot{E}$) and microphysics ($\zeta$ and $\epsilon_\mathrm{B}$) parameters for all models that fulfill the three
conditions listed above and have in addition a synchrotron spectrum that peaks in gamma-rays  with a low-energy index in the expected range ($-3/2\le\alpha\le-2/3$) show that low-values of $\zeta$ are needed to produce gamma-rays (typically $\zeta=10^{-3}-10^{-2}$), and that low values of $\epsilon_\mathrm{B}$ favor steeper low-energy photon indexes.
  
% SECTION 4.3 : A few examples of synthetic GRBs
% ---------------------------------------------------------------

\subsection{A few examples of synthetic GRBs}
In a more realistic description of the internal shock model, each pulse 
in GRBs with complex multi-pulses lightcurves
is associated with the propagation of a shock wave within the relativistic outflow \citep{daigne:00,mimica:07,mimica:10}. This propagation implies an evolution of the physical quantities in the shocked region, especially the density and therefore the magnetic field. This leads to a spectral evolution within each pulse that has already been partially described in \citet{daigne:98,daigne:03,bosnjak:09}.  
The model developed by \citet{bosnjak:09} couples a dynamical simulation of internal shock propagation within a relativistic outflow, and a detailed radiative code. This allows to predict the  lightcurves and spectral evolution in pulses for different assumptions regarding the physical conditions in the outflow.
In order to compare the results with observed
distributions of spectral parameters, we face difficulties due to several possible biases, as described in \refsec{sec:catalog}. 
To make a full comparison, one should generate noise in our synthetic bursts and take into account the response function of a given instrument before fitting the resulting spectrum by a Band function. We did not follow this procedure as our primary goal is to identify the theoretical limits for the prediction of the low-energy slope.
We computed theoretical spectra over time bins of duration $0.25$ s and measure the slope below the peak energy in the same way as in \refsec{sec:theory}.\\
 
%%%%%%%%%%%%%%%%%%%%%%%%%%%%%%%%%%%%%%%%%%%%%%%%%%%%%%%%%
% Figure 8
%
\begin{figure}[t]
\begin{center}
\includegraphics[width=0.45\textwidth]{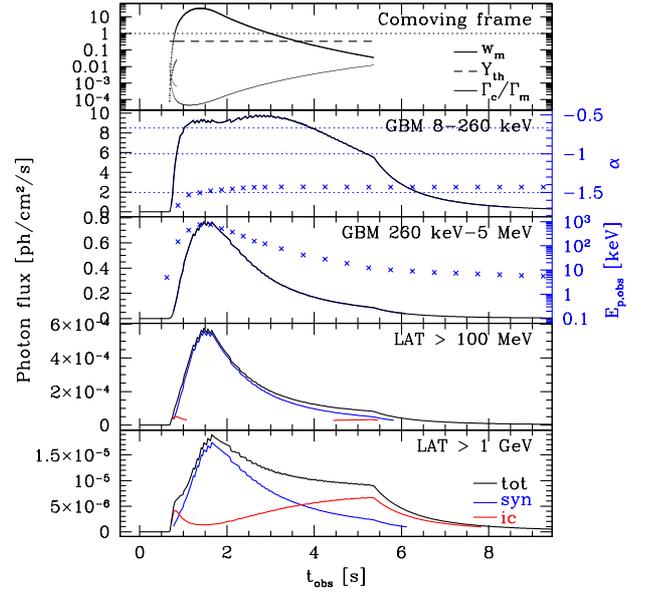}
\end{center}
\caption{\textbf{Case A : an example of a pulse generated by an internal shock with synchrotron radiation in pure fast cooling.}
The top panel plots the evolution of the three main parameters shaping the radiated spectrum (see \refsec{sec:theory}) : $w_\mathrm{m}$, $Y_\mathrm{Th}$ and $\Gamma_\mathrm{c}/\Gamma_\mathrm{m}$. The four other panels show the lightcurves in different energy channels corresponding to the GBM and the LAT on board \textit{Fermi}. The respective contributions of synchrotron and inverse Compton are also indicated. In the two GBM panels, the evolution of the low-energy slope and peak energy of the soft gamma-ray component is also plotted. In this case, the standard fast cooling synchrotron slope $\alpha=-3/2$ is found for the whole duration.}
\label{fig:exampleA}
\end{figure}
%%%%%%%%%%%%%%%%%%%%%%%%%%%%%%%%%%%%%%%%%%%%%%%%%%%%%%%%% 
 
\begin{figure}[!t]
\begin{center}
\includegraphics[width=0.45\textwidth]{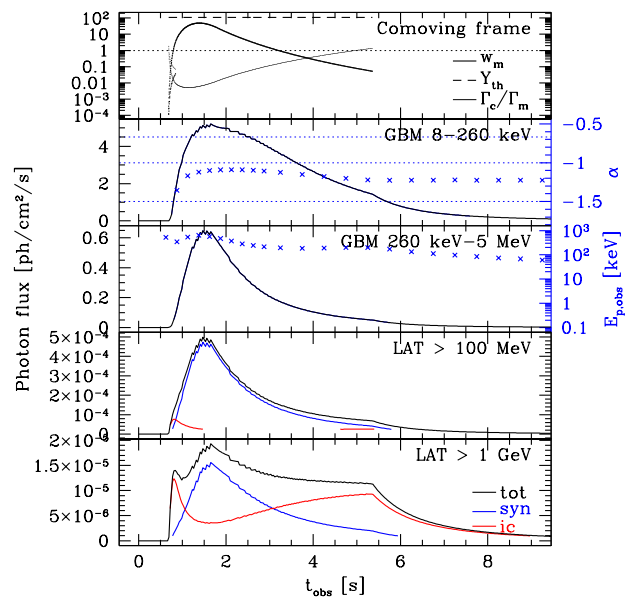}
\end{center}
\caption{\textbf{Case B : an example of a pulse generated by an internal shock with synchrotron radiation in fast cooling regime affected by non negligible inverse Compton scatterings in Klein-Nishina regime.}
Same as in \reffig{fig:exampleA}. The slope $\alpha$ is steeper ($-3/2 < \alpha < -1$ for the whole duration) than in case A, due to comparable values of $w_\mathrm{m}$ but higher values of $Y_\mathrm{Th}$ (see the theoretical interpretation in \refsec{sec:theory}).}
\label{fig:exampleB}
\end{figure}

In the examples below, we adopt the same reference case as in \citet{bosnjak:09} : a single pulse burst is generated by a relativistic ejection lasting for $t_\mathrm{w}=2\ \mathrm{s}$ with a constant $\dot{E}$ and a Lorentz factor increasing from 100 to 400 (see Figure~1 in \citealt{bosnjak:09}). 
Constant microphysics parameters are assumed during the whole evolution. This is a simplifying assumption due to our poor knowledge of the physical processes at work in mildly relativistic shocks. However, as the diversity of GRBs and their afterglows seem to indicate that these parameters are not universal, they are most probably evolving with shock conditions, which could impact the spectral evolution within a pulse \citep{daigne:03}.
We adopt here $\epsilon_\mathrm{e}=1/3$ and $p=2.5$ and adjust $\zeta$ and $\epsilon_\mathrm{B}$ to have the peak energy of the pulse well within the GBM range.
We consider the following examples to illustrate the possible range of $\alpha$ :
\begin{itemize}
\item Case A : $\dot{E}=10^{54}\ \mathrm{erg/s}$, $\epsilon_\mathrm{B}=1/3$ and $\zeta=3\times 10^{-3}$. This case is plotted in \reffig{fig:exampleA} and shows the standard slope $\alpha=-3/2$. The peak energy is $E_\mathrm{p,obs}\simeq 800\,\mathrm{keV}$ at the peak of the pulse.
\item Case B : $\dot{E}=10^{54}\ \mathrm{erg/s}$, $\epsilon_\mathrm{B}=10^{-3}$ and $\zeta=10^{-3}$. This case is plotted in \reffig{fig:exampleB} and shows a steeper $\alpha\simeq -1$. 
The peak energy is $E_\mathrm{p,obs}\simeq 700\,\mathrm{keV}$ at the peak of the pulse.
\end{itemize}
Intermediate values of $\epsilon_\mathrm{B}$ between case A and B would lead to intermediate values of $\alpha$ between $-3/2$ and $-1$. In both examples, it appears clearly that all spectral parameters are evolving during a given pulse. The evolution for the peak energy is stronger in case A than in case B, whereas the low-energy photon index $\alpha$ evolves more strongly in the second case. Note that in these examples, the direct emission from pulse ends at  $t_\mathrm{obs}\sim 5.5\,\mathrm{s}$. The flux observed after this time is due to the high latitude emission. In more complex bursts with multi-pulses lightcurves, the spectral properties should in principle be governed by the dominant pulse at a given time. The tail and high-latitude emission of a pulse can be observed only if it is followed by a period of inactivity.\\

Examples A and B are very encouraging as they illustrate that low-energy slopes $\alpha$ can be expected in the range $-3/2\le \alpha\le -1$ in the internal shock model with dominant synchrotron radiation. On the other hand, keeping the same assumption for the dynamics of the internal shocks as in case A and B, it seems difficult to find microphysics parameters leading to even steeper slopes $-1\le\alpha\le-2/3$. 
This can be understood from the two shell model presented in the previous subsection. To reach the necessary condition $\Gamma_\mathrm{c}/\Gamma_\mathrm{m}\sim 0.1-1$, it is necessary to have collisions at lower radii, and/or with larger bulk Lorentz factor, and/or to reduce the magnetic field (see \refeq{eq:estimgc}). This can be achieved in different ways : decreasing the contrast $\kappa$, increasing the variability timescale $\tau$, increasing the Lorentz factor $\bar{\Gamma}$, or reducing the kinetic energy flux $\dot{E}$.
In the following example, both the Lorentz factor and the kinetic energy flux have been changed~:
\begin{itemize}
\item Case C : the dynamics is the same as in case A and B except that the Lorentz factor has been multiplied by 3 and the kinetic energy flux reduced to $\dot{E}=5\times 10^{52}\,\mathrm{erg/s}$. The microphysics parameters are $\epsilon_\mathrm{B}=0.1$ and $\zeta=10^{-3}$. This case is plotted in \reffig{fig:exampleC} and shows low-energy slopes $\alpha$ steeper than $-1$ during the whole duration of the pulse. The peak energy is $E_\mathrm{p,obs}\sim  170\,\mathrm{keV}$ at the peak of the pulse. The radiative efficiency is still reasonably high ($\sim 60\,\%$) however the end of the evolution occurs in slow cooling regime which results in a more complex evolution of the peak energy in the tail of the pulse than in the two previous examples.
\end{itemize}
This example illustrates that the "marginally fast cooling regime'' does provide low-energy slopes $\alpha > -1$.
However, following \reffig{fig:explo1}, the conditions require a smaller radius and/or a low magnetic field. This tends to favor less energetic internal shocks.

%%%%%%%%%%%%%%%%%%%%%%%%%%%%%%%%%%%%%%%%%%%%%%%%%%%%%%%%%
% Figure 9
%
%%%%%%%%%%%%%%%%%%%%%%%%%%%%%%%%%%%%%%%%%%%%%%%%%%%%%%%%%

\subsection{High-energy emission}

\begin{figure}[!t]
\begin{center}
\includegraphics[width=0.45\textwidth]{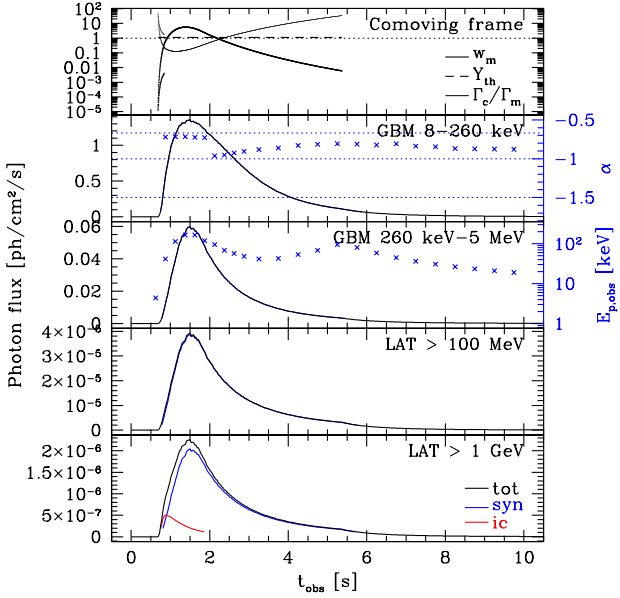}
\end{center}
\caption{\textbf{Case C : an example of a pulse generated by an internal shock with synchrotron radiation in "marginally fast cooling".}
Same as in \reffig{fig:exampleA} and~\ref{fig:exampleB}. The slope $\alpha$ is steeper ($-1 < \alpha < -2/3$ for the whole duration) than in case A and B, due to a higher ratio $\Gamma_\mathrm{c}/\Gamma_\mathrm{m}$. This ratio is of the order of $0.1-1$ at the peak of the pulse ("marginally fast cooling"), and even larger in the tail (slow cooling).}
\label{fig:exampleC}
\end{figure}

Interestingly, as already pointed out in \citet{bosnjak:09}, the scenario presented in this section -- internal shocks with dominant synchrotron radiation in the soft gamma-ray range -- require high Lorentz factors for electrons, which, because of  Klein-Nishina corrections, always limits the efficiency of inverse Compton scatterings. So the high-energy spectrum (\textit{Fermi}/LAT range) does not show any bright additional component simultaneously with the peak of the pulse in the GBM range, which seems in agreement with the GRB detection rate of \textit{Fermi}/LAT. However, as described in details in \citet{bosnjak:09}, the physical conditions in the shocked region evolve during the internal shock propagation in such a way that the $w_\mathrm{m}$ parameter decreases in the tail of the pulse (see top panel in Figs. ~\ref{fig:exampleA}--\ref{fig:exampleC}). Inverse Compton scattering progressively enters the Thomson regime and becomes more efficient, especially in the low $\epsilon_\mathrm{B}$ (high $Y_\mathrm{Th}$) case favored for steep low-energy slopes (see case B in \reffig{fig:exampleB}). This leads to the delayed emergence of an additional component in the high-energy spectrum. We will investigate in the future if this effect could explain the behaviour observed in \textit{Fermi}/LAT GRB lightcurves where delays between the GeV and the keV-MeV emission are indeed observed (see for instance GRB 080916C, \citealt{abdo:09}). 

%%%%%%%%%%%%%%%%%%%%%%%%%%%%%%%%%%%%%%%%%%%%%%%%%%%%%%%%%
% Figure 10
%
%%%%%%%%%%%%%%%%%%%%%%%%%%%%%%%%%%%%%%%%%%%%%%%%%%%%%%%%%

% =====================
% SECTION 5 : Conclusions
% =====================

\section{Discussion and conclusions}
\label{sec:conclusion}
We present here a detailed discussion of the expected value for the low-energy slope $\alpha$ of the soft gamma-ray component (BATSE -- \textit{Fermi}/GBM range) in prompt GRBs,
 in the theoretical framework where the kinetic energy of the outflow is extracted by internal shocks, and eventually injected in shock-accelerated
electrons that radiate dominantly by synchrotron radiation. Our approach is based on the model developed in \citet{bosnjak:09}, which takes into account both a full treatement of the dynamics of internal shocks and a detailed radiative calculation.

\begin{enumerate}
\item We show that in a large region of the parameters space of the internal shock model, the physical conditions in the emitting regions allow a combination of synchrotron radiation peaking in the soft gamma-ray range, and moderately efficient inverse Compton scatterings in Klein-Nishina regime. Interestingly, these scatterings affect the spectral shape of the synchrotron component, due to a better efficiency for low frequency photons. This results in a steepening of the low-energy photon index $\alpha$, with $\alpha\to -1$  \citep{derishev:01,bosnjak:09,nakar:09}.
For a large range of parameters regarding the dynamics of internal shocks (variability timescale, bulk Lorentz factor, amplitude of fluctuations of the Lorentz factor, kinetic energy flux), we produce synthetic pulses with peak energies of a few hundred keV in the observer frame, and steep slopes in the range $-3/2 \le \alpha \le -1$, at the peak of the lightcurve.  
The examples presented in the paper not only show high peak energies and steep slopes at the peak of the lightcurve, but also a general hard-to-soft spectral evolution, in agreement with observations.
This scenario constrains the microphysics in mildly relativistic shocks (shock acceleration and magnetic field amplification) : a large ($\epsilon_\mathrm{e}\sim 0.1$--$1/3$) fraction of the dissipated energy should be preferentially injected into a small ($\zeta \la 0.01$) fraction of electrons to produce a non-thermal population with high Lorentz factors, and the fraction of the energy which is injected in the magnetic field should remain low ($\epsilon_\mathrm{B}\la 10^{-3}$) to favor inverse Compton scatterings. The current knowledge of the microphysics  in mildly relativistic shocks is unfortunately still rather poor  and does not allow a comparison of these constraints with some theoretical expectations. The prediction that only a small fraction $\zeta$ of electrons are injected into a non-thermal power-law distribution leads naturally to a new question that we plan to investigate in the future : does the remaining quasi-thermal population of electrons  contribute to the emission ?

\item We also identify a regime of marginally fast-cooling synchrotron radiation with $\Gamma_\mathrm{c}\la\Gamma_\mathrm{m}$ which leads to even steeper slopes $\alpha\to -2/3$ without reducing significantly the radiative efficiency of the process ($f_\mathrm{rad}\ga 0.5$). We present an example of a synthetic pulse in this regime, with a slope $\alpha > -1$ for its whole duration. 
This requires low radii, and/or large bulk Lorentz factors, and/or weak magnetic fields.
\end{enumerate}

The present study shows that for a large region of the parameter space, internal shocks lead to spectra dominated by a bright synchrotron component in the soft gamma-ray range, with a steep slope low-energy photon-index $-3/2\le \alpha \le -1$, as observed in most prompt GRB spectra.  In this scenario, the additional Inverse Compton component at high energy is only sub-dominant and its intensity is not correlated to the intensity of the soft gamma-ray component. This seems in better agreement with \textit{Fermi}/LAT observations and GRB detection rate than other scenario, such as standard SSC in Thomson regime, where bright components are predicted at high energy.\\

On the other hand, even if steeper slopes in the range $-1\le \alpha \le -2/3$ can be achieved in the "fast marginally fast cooling regime", the corresponding constraints on the parameters of internal shocks are very strict. This regime could work  if these steep slopes are associated, on average, with weaker intensities in the light curve. Indeed, 
the marginally fast cooling regime leads not only  to a moderate radiative efficiency $f_\mathrm{rad}\simeq 0.5-0.7$ (compared to $f_\mathrm{rad}\to 1$ in the standard case), but also tends to occur in less energetic collisions. This limitation implies that this regime is probably not a robust explanation for all spectra with such steep slopes.\\

The scenario proposed in this paper can never produce low-energy slopes $\alpha$ steeper than $-2/3$, whereas such slopes are measured in a non negligible fraction of GRB spectra \citep{ryde:04,ghirlanda:03}. About 20 \%
of GRBs have more than 50 \%
of their spectra with such very steep slopes.  One possibility in such cases would be the appearance of a quasi-thermal component of photospheric origin \citep{meszaros:00,daigne:02,peer:08,beloborodov:10,peer:10}. The emission from both the photosphere and internal shocks
have a similar duration, the latter having only a very short lag behind the first. The intensity of the photospheric 
emission
depends strongly on the unknown mechanism responsible for the acceleration of the relativistic outflow. A pure fireball would lead to a dominant thermal emission, whereas mechanisms such as the "magnetic rocket" recently suggested by \citet{granot:10b} would on the other hand correspond to much colder jets with only a faint photospheric emission. In addition, the way it will superimpose on the non-thermal emission from internal shocks depends on the details of the initial distribution of the Lorentz factor in the flow \citep[see][]{daigne:02}. Very steep slopes could be observed
in the time bins where the photospheric emission is dominant.\\

Other effects could be a source of additional complexity in the prompt GRB spectrum within the scenario proposed in this paper.
As it is required that only a fraction of available electrons are shock accelerated into a non-thermal power-law distribution, possible extra components in the spectrum could be associated with the remaining thermal population of electrons. Preliminary investigations show that the synchrotron radiation from these electrons is easily self-absorbed, in agreement with \citet{zou:09}. On the other hand, due to lower electron Lorentz factors, inverse Compton scatterings by these electrons usually occur in Thomson regime, which is efficient. This could lead to additional components at low and/or high energy that offer interesting perspectives for the interpretation of the complex behaviour observed in GRBs detected by \textit{Fermi}/LAT. We also find that in some cases, a non negligible fraction of the radiated energy is re-injected in pairs due to $\gamma\gamma$ annihilation. These pairs can radiate and scatter photons as well. These second order effects are not included in the present version of the model and could impact the final spectra shape.

\begin{acknowledgements}
The authors thank R. Mochkovitch, A. Pe'er, P. Kumar and E. Nakar for valuable discussions on this work. 
The authors thank Y. Kaneko for her help regarding BATSE GRB spectral results.
The authors acknowledge support from the Agence Nationale de la Recherche via contract ANR--JC05--44822. 
Z.B. and F.D. acknowledge the French Space Agency (CNES) for financial support. 
G.D. acknowledges support from the European Community via contract ERC--StG--200911.
\end{acknowledgements}

\bibliographystyle{aa}
\bibliography{paper}

\begin{thebibliography}{74}
\expandafter\ifx\csname natexlab\endcsname\relax\def\natexlab#1{#1}\fi

\bibitem[{{Abdo} {et~al.}(2009){Abdo}, {Ackermann}, {Arimoto}, {Asano},
  {Atwood}, {Axelsson}, {Baldini}, {Ballet}, {Band}, {Barbiellini}, {Baring},
  {Bastieri}, {Battelino}, {Baughman}, {Bechtol}, {Bellardi}, {Bellazzini},
  {Berenji}, {Bhat}, {Bissaldi}, {Blandford}, {Bloom}, {Bogaert}, {Bogart},
  {Bonamente}, {Bonnell}, {Borgland}, {Bouvier}, {Bregeon}, {Brez}, {Briggs},
  {Brigida}, {Bruel}, {Burnett}, {Burrows}, {Busetto}, {Caliandro}, {Cameron},
  {Caraveo}, {Casandjian}, {Ceccanti}, {Cecchi}, {Celotti}, {Charles},
  {Chekhtman}, {Cheung}, {Chiang}, {Ciprini}, {Claus}, {Cohen-Tanugi},
  {Cominsky}, {Connaughton}, {Conrad}, {Costamante}, {Cutini}, {DeKlotz},
  {Dermer}, {de Angelis}, {de Palma}, {Digel}, {Dingus}, {do Couto e Silva},
  {Drell}, {Dubois}, {Dumora}, {Edmonds}, {Evans}, {Fabiani}, {Farnier},
  {Favuzzi}, {Finke}, {Fishman}, {Focke}, {Frailis}, {Fukazawa}, {Funk},
  {Fusco}, {Gargano}, {Gasparrini}, {Gehrels}, {Germani}, {Giebels},
  {Giglietto}, {Giommi}, {Giordano}, {Glanzman}, {Godfrey}, {Goldstein},
  {Granot}, {Greiner}, {Grenier}, {Grondin}, {Grove}, {Guillemot}, {Guiriec},
  {Haller}, {Hanabata}, {Harding}, {Hayashida}, {Hays}, {Morata}, {Hoover},
  {Hughes}, {J{\'o}hannesson}, {Johnson}, {Johnson}, {Johnson}, {Johnson},
  {Kamae}, {Katagiri}, {Kataoka}, {Kavelaars}, {Kawai}, {Kelly}, {Kennea},
  {Kerr}, {Kippen}, {Kn{\"o}dlseder}, {Kocevski}, {Kocian}, {Komin},
  {Kouveliotou}, {Kuehn}, {Kuss}, {Lande}, {Landriu}, {Larsson}, {Latronico},
  {Lavalley}, {Lee}, {Lee}, {Lemoine-Goumard}, {Lichti}, {Longo}, {Loparco},
  {Lott}, {Lovellette}, {Lubrano}, {Madejski}, {Makeev}, {Marangelli},
  {Mazziotta}, {McBreen}, {McEnery}, {McGlynn}, {Meegan}, {M{\'e}sz{\'a}ros},
  {Meurer}, {Michelson}, {Minuti}, {Mirizzi}, {Mitthumsiri}, {Mizuno},
  {Moiseev}, {Monte}, {Monzani}, {Moretti}, {Morselli}, {Moskalenko}, {Murgia},
  {Nakamori}, {Nelson}, {Nolan}, {Norris}, {Nuss}, {Ohno}, {Ohsugi}, {Okumura},
  {Omodei}, {Orlando}, {Ormes}, {Ozaki}, {Paciesas}, {Paneque}, {Panetta},
  {Parent}, {Pelassa}, {Pepe}, {Perri}, {Pesce-Rollins}, {Petrosian},
  {Pinchera}, {Piron}, {Porter}, {Preece}, {Rain{\`o}}, {Ramirez-Ruiz},
  {Rando}, {Rapposelli}, {Razzano}, {Razzaque}, {Rea}, {Reimer}, {Reimer},
  {Reposeur}, {Reyes}, {Ritz}, {Rochester}, {Rodriguez}, {Roth}, {Ryde},
  {Sadrozinski}, {Sanchez}, {Sander}, {Parkinson}, {Scargle}, {Schalk},
  {Segal}, {Sgr{\`o}}, {Shimokawabe}, {Siskind}, {Smith}, {Smith}, {Spandre},
  {Spinelli}, {Stamatikos}, {Starck}, {Stecker}, {Steinle}, {Stephens},
  {Strickman}, {Suson}, {Tagliaferri}, {Tajima}, {Takahashi}, {Takahashi},
  {Tanaka}, {Tenze}, {Thayer}, {Thayer}, {Thompson}, {Tibaldo}, {Torres},
  {Tosti}, {Tramacere}, {Turri}, {Tuvi}, {Usher}, {van der Horst}, {Vigiani},
  {Vilchez}, {Vitale}, {von Kienlin}, {Waite}, {Williams}, {Wilson-Hodge},
  {Winer}, {Wood}, {Wu}, {Yamazaki}, {Ylinen}, {Ziegler}, {The Fermi LAT
  Collaboration}, \& {The Fermi GBM Collaboration}}]{abdo:09}
{Abdo}, A.~A., {Ackermann}, M., {Arimoto}, M., {et~al.} 2009, Science, 323,
  1688

\bibitem[{{Asano} \& {Inoue}(2007)}]{asano:07}
{Asano}, K. \& {Inoue}, S. 2007, \apj, 671, 645

\bibitem[{{Asano} \& {Terasawa}(2009)}]{asano:09}
{Asano}, K. \& {Terasawa}, T. 2009, \apj, 705, 1714

\bibitem[{{Band} {et~al.}(1993){Band}, {Matteson}, {Ford}, {Schaefer},
  {Palmer}, {Teegarden}, {Cline}, {Briggs}, {Paciesas}, {Pendleton}, {Fishman},
  {Kouveliotou}, {Meegan}, {Wilson}, \& {Lestrade}}]{band:93}
{Band}, D., {Matteson}, J., {Ford}, L., {et~al.} 1993, \apj, 413, 281

\bibitem[{{Baring} \& {Braby}(2004)}]{baring:04}
{Baring}, M.~G. \& {Braby}, M.~L. 2004, \apj, 613, 460

\bibitem[{{Barraud} {et~al.}(2005){Barraud}, {Daigne}, {Mochkovitch}, \&
  {Atteia}}]{barraud:05}
{Barraud}, C., {Daigne}, F., {Mochkovitch}, R., \& {Atteia}, J.~L. 2005, \aap,
  440, 809

\bibitem[{{Belmont} {et~al.}(2008){Belmont}, {Malzac}, \&
  {Marcowith}}]{belmont:08}
{Belmont}, R., {Malzac}, J., \& {Marcowith}, A. 2008, \aap, 491, 617

\bibitem[{{Beloborodov}(2010)}]{beloborodov:10}
{Beloborodov}, A.~M. 2010, \mnras, 407, 1033

\bibitem[{{Borgonovo} \& {Ryde}(2001)}]{borgonovo:01}
{Borgonovo}, L. \& {Ryde}, F. 2001, \apj, 548, 770

\bibitem[{{Bo{\v s}njak} {et~al.}(2009){Bo{\v s}njak}, {Daigne}, \&
  {Dubus}}]{bosnjak:09}
{Bo{\v s}njak}, {\v Z}., {Daigne}, F., \& {Dubus}, G. 2009, \aap, 498, 677

\bibitem[{{Crider} {et~al.}(1997){Crider}, {Liang}, {Smith}, {Preece},
  {Briggs}, {Pendleton}, {Paciesas}, {Band}, \& {Matteson}}]{crider:97}
{Crider}, A., {Liang}, E.~P., {Smith}, I.~A., {et~al.} 1997, \apjl, 479, L39+

\bibitem[{{Daigne} \& {Mochkovitch}(1998)}]{daigne:98}
{Daigne}, F. \& {Mochkovitch}, R. 1998, \mnras, 296, 275

\bibitem[{{Daigne} \& {Mochkovitch}(2000)}]{daigne:00}
{Daigne}, F. \& {Mochkovitch}, R. 2000, \aap, 358, 1157

\bibitem[{{Daigne} \& {Mochkovitch}(2002)}]{daigne:02}
{Daigne}, F. \& {Mochkovitch}, R. 2002, \mnras, 336, 1271

\bibitem[{{Daigne} \& {Mochkovitch}(2003)}]{daigne:03}
{Daigne}, F. \& {Mochkovitch}, R. 2003, \mnras, 342, 587

\bibitem[{{Derishev} {et~al.}(2001){Derishev}, {Kocharovsky}, \&
  {Kocharovsky}}]{derishev:01}
{Derishev}, E.~V., {Kocharovsky}, V.~V., \& {Kocharovsky}, V.~V. 2001, \aap,
  372, 1071

\bibitem[{{Drenkhahn} \& {Spruit}(2002)}]{drenkhahn:02}
{Drenkhahn}, G. \& {Spruit}, H.~C. 2002, \aap, 391, 1141

\bibitem[{{Ford} {et~al.}(1995){Ford}, {Band}, {Matteson}, {Briggs},
  {Pendleton}, {Preece}, {Paciesas}, {Teegarden}, {Palmer}, {Schaefer},
  {Cline}, {Fishman}, {Kouveliotou}, {Meegan}, {Wilson}, \&
  {Lestrade}}]{ford:95}
{Ford}, L.~A., {Band}, D.~L., {Matteson}, J.~L., {et~al.} 1995, \apj, 439, 307

\bibitem[{{Ghirlanda} {et~al.}(2003){Ghirlanda}, {Celotti}, \&
  {Ghisellini}}]{ghirlanda:03}
{Ghirlanda}, G., {Celotti}, A., \& {Ghisellini}, G. 2003, \aap, 406, 879

\bibitem[{{Ghirlanda} {et~al.}(2010){Ghirlanda}, {Nava}, \&
  {Ghisellini}}]{ghirlanda:10}
{Ghirlanda}, G., {Nava}, L., \& {Ghisellini}, G. 2010, \aap, 511, A43+

\bibitem[{{Ghisellini} \& {Celotti}(1999)}]{ghisellini:99}
{Ghisellini}, G. \& {Celotti}, A. 1999, \aaps, 138, 527

\bibitem[{{Ghisellini} {et~al.}(2000){Ghisellini}, {Celotti}, \&
  {Lazzati}}]{ghisellini:00}
{Ghisellini}, G., {Celotti}, A., \& {Lazzati}, D. 2000, \mnras, 313, L1

\bibitem[{{Giannios} \& {Spitkovsky}(2009)}]{giannios:09}
{Giannios}, D. \& {Spitkovsky}, A. 2009, \mnras, 400, 330

\bibitem[{{Giannios} \& {Spruit}(2005)}]{giannios:05}
{Giannios}, D. \& {Spruit}, H.~C. 2005, \aap, 430, 1

\bibitem[{{Giannios} \& {Spruit}(2007)}]{giannios:07}
{Giannios}, D. \& {Spruit}, H.~C. 2007, \aap, 469, 1

\bibitem[{{Granot} {et~al.}(2008){Granot}, {Cohen-Tanugi}, \& {do Couto e
  Silva}}]{granot:08}
{Granot}, J., {Cohen-Tanugi}, J., \& {do Couto e Silva}, E. 2008, \apj, 677, 92

\bibitem[{{Granot} {et~al.}(2010{\natexlab{a}}){Granot}, {for the Fermi LAT
  Collaboration}, \& {the GBM Collaboration}}]{granot:10a}
{Granot}, J., {for the Fermi LAT Collaboration}, \& {the GBM Collaboration}.
  2010{\natexlab{a}}, in Italian Physical Society, Vol. 102, The Shocking
  Universe - Gamma-Ray Bursts and High Energy Shock phenomena, ed. {G.
  Chincarini, P. d'Avanzo, R. Margutti \& R. Salvaterra}, 177--190,
  arXiv:1003.2452

\bibitem[{{Granot} {et~al.}(2010{\natexlab{b}}){Granot}, {Komissarov}, \&
  {Spitkovsky}}]{granot:10b}
{Granot}, J., {Komissarov}, S., \& {Spitkovsky}, A. 2010{\natexlab{b}}, ArXiv
  e-prints

\bibitem[{{Heise} {et~al.}(2001){Heise}, {in't Zand}, {Kippen}, \&
  {Woods}}]{heise:01}
{Heise}, J., {in't Zand}, J., {Kippen}, R.~M., \& {Woods}, P.~M. 2001, in
  Gamma-ray Bursts in the Afterglow Era, ed. {E.~Costa, F.~Frontera, \&
  J.~Hjorth}, 16--+

\bibitem[{{Kaneko} {et~al.}(2006){Kaneko}, {Preece}, {Briggs}, {Paciesas},
  {Meegan}, \& {Band}}]{kaneko:06}
{Kaneko}, Y., {Preece}, R.~D., {Briggs}, M.~S., {et~al.} 2006, \apjs, 166, 298

\bibitem[{{Kobayashi} {et~al.}(1997){Kobayashi}, {Piran}, \&
  {Sari}}]{kobayashi:97}
{Kobayashi}, S., {Piran}, T., \& {Sari}, R. 1997, \apj, 490, 92

\bibitem[{{Krimm} {et~al.}(2009){Krimm}, {Yamaoka}, {Sugita}, {Ohno},
  {Sakamoto}, {Barthelmy}, {Gehrels}, {Hara}, {Norris}, {Ohmori}, {Onda},
  {Sato}, {Tanaka}, {Tashiro}, \& {Yamauchi}}]{krimm:09}
{Krimm}, H.~A., {Yamaoka}, K., {Sugita}, S., {et~al.} 2009, \apj, 704, 1405

\bibitem[{{Kumar} \& {McMahon}(2008)}]{kumar:08}
{Kumar}, P. \& {McMahon}, E. 2008, \mnras, 384, 33

\bibitem[{{Liang} {et~al.}(1997){Liang}, {Kusunose}, {Smith}, \&
  {Crider}}]{liang:97}
{Liang}, E., {Kusunose}, M., {Smith}, I.~A., \& {Crider}, A. 1997, \apjl, 479,
  L35+

\bibitem[{{Lithwick} \& {Sari}(2001)}]{lithwick:01}
{Lithwick}, Y. \& {Sari}, R. 2001, \apj, 555, 540

\bibitem[{{Lloyd} \& {Petrosian}(2000)}]{lloyd:00}
{Lloyd}, N.~M. \& {Petrosian}, V. 2000, \apj, 543, 722

\bibitem[{{Lloyd-Ronning} \& {Petrosian}(2002)}]{lloyd-ronning:02}
{Lloyd-Ronning}, N.~M. \& {Petrosian}, V. 2002, \apj, 565, 182

\bibitem[{{Lyutikov} \& {Blandford}(2003)}]{lyutikov:03}
{Lyutikov}, M. \& {Blandford}, R. 2003, arXiv:astro-ph/0312347

\bibitem[{{Mallozzi} {et~al.}(1995){Mallozzi}, {Paciesas}, {Pendleton},
  {Briggs}, {Preece}, {Meegan}, \& {Fishman}}]{mallozzi:95}
{Mallozzi}, R.~S., {Paciesas}, W.~S., {Pendleton}, G.~N., {et~al.} 1995, \apj,
  454, 597

\bibitem[{{Martins} {et~al.}(2009){Martins}, {Fonseca}, {Silva}, \&
  {Mori}}]{martins:09}
{Martins}, S.~F., {Fonseca}, R.~A., {Silva}, L.~O., \& {Mori}, W.~B. 2009,
  \apjl, 695, L189

\bibitem[{{Medvedev}(2000)}]{medvedev:00}
{Medvedev}, M.~V. 2000, \apj, 540, 704

\bibitem[{{Meszaros} \& {Rees}(1997)}]{meszaros:97}
{Meszaros}, P. \& {Rees}, M.~J. 1997, \apjl, 482, L29+

\bibitem[{{M{\'e}sz{\'a}ros} \& {Rees}(2000)}]{meszaros:00}
{M{\'e}sz{\'a}ros}, P. \& {Rees}, M.~J. 2000, \apj, 530, 292

\bibitem[{{Mimica} \& {Aloy}(2010)}]{mimica:10}
{Mimica}, P. \& {Aloy}, M.~A. 2010, \mnras, 401, 525

\bibitem[{{Mimica} {et~al.}(2007){Mimica}, {Aloy}, \& {M{\"u}ller}}]{mimica:07}
{Mimica}, P., {Aloy}, M.~A., \& {M{\"u}ller}, E. 2007, \aap, 466, 93

\bibitem[{{Nakar} {et~al.}(2009){Nakar}, {Ando}, \& {Sari}}]{nakar:09}
{Nakar}, E., {Ando}, S., \& {Sari}, R. 2009, \apj, 703, 675

\bibitem[{{Omodei} {et~al.}(2009){Omodei}, {for the Fermi LAT}, \& {Fermi GBM
  collaborations}}]{omodei:09}
{Omodei}, N., {for the Fermi LAT}, \& {Fermi GBM collaborations}. 2009,
  arXiv:0907.0715

\bibitem[{{Panaitescu} \& {M{\'e}sz{\'a}ros}(2000)}]{panaitescu:00}
{Panaitescu}, A. \& {M{\'e}sz{\'a}ros}, P. 2000, \apjl, 544, L17

\bibitem[{{Pe'er}(2008)}]{peer:08}
{Pe'er}, A. 2008, \apj, 682, 463

\bibitem[{{Pe'er} \& {Waxman}(2005)}]{peer:05}
{Pe'er}, A. \& {Waxman}, E. 2005, \apj, 628, 857

\bibitem[{{Pe'er} \& {Zhang}(2006)}]{peer:06}
{Pe'er}, A. \& {Zhang}, B. 2006, \apj, 653, 454

\bibitem[{{Pe'er} {et~al.}(2010){Pe'er}, {Zhang}, {Ryde}, {McGlynn}, {Zhang},
  {Preece}, \& {Kouveliotou}}]{peer:10}
{Pe'er}, A., {Zhang}, B., {Ryde}, F., {et~al.} 2010, arXiv:1007.2228

\bibitem[{{P{\'e}langeon} {et~al.}(2008){P{\'e}langeon}, {Atteia}, {Nakagawa},
  {Hurley}, {Yoshida}, {Vanderspek}, {Suzuki}, {Kawai}, {Pizzichini},
  {Bo{\"e}r}, {Braga}, {Crew}, {Donaghy}, {Dezalay}, {Doty}, {Fenimore},
  {Galassi}, {Graziani}, {Jernigan}, {Lamb}, {Levine}, {Manchanda}, {Martel},
  {Matsuoka}, {Olive}, {Prigozhin}, {Ricker}, {Sakamoto}, {Shirasaki},
  {Sugita}, {Takagishi}, {Tamagawa}, {Villasenor}, {Woosley}, \&
  {Yamauchi}}]{pelangeon:08}
{P{\'e}langeon}, A., {Atteia}, J., {Nakagawa}, Y.~E., {et~al.} 2008, \aap, 491,
  157

\bibitem[{{Piran} {et~al.}(2009){Piran}, {Sari}, \& {Zou}}]{piran:09}
{Piran}, T., {Sari}, R., \& {Zou}, Y. 2009, \mnras, 393, 1107

\bibitem[{{Preece} {et~al.}(1998){Preece}, {Briggs}, {Mallozzi}, {Pendleton},
  {Paciesas}, \& {Band}}]{preece:98}
{Preece}, R.~D., {Briggs}, M.~S., {Mallozzi}, R.~S., {et~al.} 1998, \apjl, 506,
  L23

\bibitem[{{Preece} {et~al.}(2000){Preece}, {Briggs}, {Mallozzi}, {Pendleton},
  {Paciesas}, \& {Band}}]{preece:00}
{Preece}, R.~D., {Briggs}, M.~S., {Mallozzi}, R.~S., {et~al.} 2000, \apjs, 126,
  19

\bibitem[{{Racusin} {et~al.}(2008){Racusin}, {Karpov}, {Sokolowski}, {Granot},
  {Wu}, {Pal'Shin}, {Covino}, {van der Horst}, {Oates}, {Schady}, {Smith},
  {Cummings}, {Starling}, {Piotrowski}, {Zhang}, {Evans}, {Holland}, {Malek},
  {Page}, {Vetere}, {Margutti}, {Guidorzi}, {Kamble}, {Curran}, {Beardmore},
  {Kouveliotou}, {Mankiewicz}, {Melandri}, {O'Brien}, {Page}, {Piran},
  {Tanvir}, {Wrochna}, {Aptekar}, {Barthelmy}, {Bartolini}, {Beskin}, {Bondar},
  {Bremer}, {Campana}, {Castro-Tirado}, {Cucchiara}, {Cwiok}, {D'Avanzo},
  {D'Elia}, {Della Valle}, {de Ugarte Postigo}, {Dominik}, {Falcone}, {Fiore},
  {Fox}, {Frederiks}, {Fruchter}, {Fugazza}, {Garrett}, {Gehrels},
  {Golenetskii}, {Gomboc}, {Gorosabel}, {Greco}, {Guarnieri}, {Immler},
  {Jelinek}, {Kasprowicz}, {La Parola}, {Levan}, {Mangano}, {Mazets},
  {Molinari}, {Moretti}, {Nawrocki}, {Oleynik}, {Osborne}, {Pagani}, {Pandey},
  {Paragi}, {Perri}, {Piccioni}, {Ramirez-Ruiz}, {Roming}, {Steele}, {Strom},
  {Testa}, {Tosti}, {Ulanov}, {Wiersema}, {Wijers}, {Winters}, {Zarnecki},
  {Zerbi}, {M{\'e}sz{\'a}ros}, {Chincarini}, \& {Burrows}}]{racusin:08}
{Racusin}, J.~L., {Karpov}, S.~V., {Sokolowski}, M., {et~al.} 2008, \nat, 455,
  183

\bibitem[{{Ramirez-Ruiz} \& {Fenimore}(2000)}]{ramirezruiz:00}
{Ramirez-Ruiz}, E. \& {Fenimore}, E.~E. 2000, \apj, 539, 712

\bibitem[{{Rees}(1967)}]{rees:67}
{Rees}, M.~J. 1967, \mnras, 137, 429

\bibitem[{{Rees} \& {M\'{e}sz\'{a}ros}(1994)}]{rees:94}
{Rees}, M.~J. \& {M\'{e}sz\'{a}ros}, P. 1994, \apjl, 430, L93

\bibitem[{{Ryde}(2004)}]{ryde:04}
{Ryde}, F. 2004, \apj, 614, 827

\bibitem[{{Sakamoto} {et~al.}(2008){Sakamoto}, {Hullinger}, {Sato}, {Yamazaki},
  {Barbier}, {Barthelmy}, {Cummings}, {Fenimore}, {Gehrels}, {Krimm}, {Lamb},
  {Markwardt}, {Osborne}, {Palmer}, {Parsons}, {Stamatikos}, \&
  {Tueller}}]{sakamoto:08}
{Sakamoto}, T., {Hullinger}, D., {Sato}, G., {et~al.} 2008, \apj, 679, 570

\bibitem[{{Sakamoto} {et~al.}(2005){Sakamoto}, {Lamb}, {Kawai}, {Yoshida},
  {Graziani}, {Fenimore}, {Donaghy}, {Matsuoka}, {Suzuki}, {Ricker}, {Atteia},
  {Shirasaki}, {Tamagawa}, {Torii}, {Galassi}, {Doty}, {Vanderspek}, {Crew},
  {Villasenor}, {Butler}, {Prigozhin}, {Jernigan}, {Barraud}, {Boer},
  {Dezalay}, {Olive}, {Hurley}, {Levine}, {Monnelly}, {Martel}, {Morgan},
  {Woosley}, {Cline}, {Braga}, {Manchanda}, {Pizzichini}, {Takagishi}, \&
  {Yamauchi}}]{sakamoto:05}
{Sakamoto}, T., {Lamb}, D.~Q., {Kawai}, N., {et~al.} 2005, \apj, 629, 311

\bibitem[{{Sari} \& {Esin}(2001)}]{sari:01}
{Sari}, R. \& {Esin}, A.~A. 2001, \apj, 548, 787

\bibitem[{{Sari} {et~al.}(1996){Sari}, {Narayan}, \& {Piran}}]{sari:96}
{Sari}, R., {Narayan}, R., \& {Piran}, T. 1996, \apj, 473, 204

\bibitem[{{Sari} {et~al.}(1998){Sari}, {Piran}, \& {Narayan}}]{sari:98}
{Sari}, R., {Piran}, T., \& {Narayan}, R. 1998, \apjl, 497, L17

\bibitem[{{Spitkovsky}(2008{\natexlab{a}})}]{spitkovsky:08b}
{Spitkovsky}, A. 2008{\natexlab{a}}, \apjl, 673, L39

\bibitem[{{Spitkovsky}(2008{\natexlab{b}})}]{spitkovsky:08a}
{Spitkovsky}, A. 2008{\natexlab{b}}, \apjl, 682, L5

\bibitem[{{Spruit} {et~al.}(2001){Spruit}, {Daigne}, \&
  {Drenkhahn}}]{spruit:01}
{Spruit}, H.~C., {Daigne}, F., \& {Drenkhahn}, G. 2001, \aap, 369, 694

\bibitem[{{Stern} \& {Poutanen}(2004)}]{stern:04}
{Stern}, B.~E. \& {Poutanen}, J. 2004, \mnras, 352, L35

\bibitem[{{Thompson}(1994)}]{thompson:94}
{Thompson}, C. 1994, \mnras, 270, 480

\bibitem[{{Vurm} \& {Poutanen}(2009)}]{vurm:09}
{Vurm}, I. \& {Poutanen}, J. 2009, \apj, 698, 293

\bibitem[{{Wang} {et~al.}(2009){Wang}, {Li}, {Dai}, \&
  {M{\'e}sz{\'a}ros}}]{wang:09}
{Wang}, X., {Li}, Z., {Dai}, Z., \& {M{\'e}sz{\'a}ros}, P. 2009, \apjl, 698,
  L98

\bibitem[{{Zou} {et~al.}(2009){Zou}, {Piran}, \& {Sari}}]{zou:09}
{Zou}, Y., {Piran}, T., \& {Sari}, R. 2009, \apjl, 692, L92

\end{thebibliography}
\end{document}